\documentclass[reprint, onecolumn, aps, amsmath, amssymb,prd]{revtex4-2}
\usepackage{verbatim}
\usepackage{bm}
\usepackage{xcolor}
\usepackage{fullpage}
\usepackage{graphicx}
\usepackage{natbib}
\usepackage{hyperref}

\newcommand{\be}{\begin{equation}}
\newcommand{\ee}{\end{equation}}
\newcommand{\bea}{\begin{eqnarray}}
\newcommand{\eea}{\end{eqnarray}}
\def\ba#1\ea{\begin{align}#1\end{align}}

\def\({\left(}
\def\){\right)}

\def\[{\left[}
\def\]{\right]}

\newcommand{\refeq}[1]{Eq.~(\ref{eq:#1})}
\newcommand{\refeqs}[2]{Eqs.~(\ref{eq:#1})--(\ref{eq:#2})}
\newcommand{\reffig}[1]{Fig.~\ref{fig:#1}}
\newcommand{\reffigs}[2]{Figs.~(\ref{fig:#1})--(\ref{fig:#2})}
\newcommand{\refsec}[1]{Sec.~\ref{sec:#1}}
\newcommand{\refapp}[1]{App.~\ref{app:#1}}
\newcommand{\reftab}[1]{Tab.~\ref{tab:#1}}
\newcommand{\vs}{\nonumber\\}

\newcommand{\tj}[6]{ \begin{pmatrix}
   #1 & #2 & #3 \\
   #4 & #5 & #6 
\end{pmatrix}}

\newcommand{\sj}[6]{ \begin{Bmatrix}
   #1 & #2 & #3 \\
   #4 & #5 & #6 
\end{Bmatrix}}

\def\nhat{\hat{\bm{n}}}
\def\khat{\hat{\bm{k}}}

\def\qhat{\hat{\bm{q}}}
\def\rhat{\hat{\bm{r}}}
\def\phat{\hat{\bm{p}}}

\begin{document}
\definecolor{RedWine}{rgb}{0.743,0,0}
\definecolor{GrassGreen}{rgb}{0.125,0.75,0.125}
\definecolor{RoyalBlue}{rgb}{0.25,0.41,0.88}
\renewcommand{\dj}[1]{\textcolor{RedWine}{\textbf{[DJ: #1]}}}
\newcommand{\jt}[1]{\textcolor{GrassGreen}{\textbf{[JT: #1]}}}
\newcommand{\hg}[1]{\textcolor{RoyalBlue}{\textbf{[HG: #1]}}}

\title{Fast calculation of nonlinear redshift-space galaxy power spectrum including selection bias}

\author{Joseph Tomlinson}
\email{jxt732@psu.edu}
\affiliation{Department of Astronomy and Astrophysics and Institute for Gravitation and the Cosmos, \\
The Pennsylvania State University, University Park, PA 16802, USA}

\author{Henry S. Grasshorn Gebhardt}
\email{hsg113@psu.edu}
\thanks{now NASA Postdoctoral Fellow at Jet Propulsion Laboratory, 4800 Oak Grove Drive, Pasadena, CA 91109}
\affiliation{Department of Astronomy and Astrophysics and Institute for Gravitation and the Cosmos, \\
The Pennsylvania State University, University Park, PA 16802, USA}

\author{Donghui Jeong}
\email{djeong@psu.edu}
\affiliation{Department of Astronomy and Astrophysics and Institute for Gravitation and the Cosmos, \\
The Pennsylvania State University, University Park, PA 16802, USA}

\begin{abstract}
We present a fast implementation of the next-to-leading 
order (1-loop) redshift-space galaxy power spectrum by 
using FFTlog-based methods.
[V.  Desjacques,  D.  Jeong,  and  F.  Schmidt,  JCAP
\textbf{1812} (12), 035] have shown that the 1-loop galaxy
power spectrum in redshift space can be computed with 28 
independent loop integrals with 22 bias parameters.
Analytical calculation of the angular part of the loop 
integrals leaves the radial part in the form of a spherical
Bessel transformation that is ready to be integrated
numerically by using the FFTLog transformation. We find 
that the original 28 loop integrals can be solved with a 
total of 85 unique FFTLog transformations, yet leading to a 
few orders of magnitude speed up over traditional
multi-dimensional integration. The code used in this work is publicly available at 
\url{https://github.com/JosephTomlinson/GeneralBiasPk}
\end{abstract}
\maketitle

\section{Introduction} \label{sec:intro}
The next frontier of precision cosmology is the study of Large Scale Structure (LSS) traced by the distribution of galaxies with many LSS surveys imminent; for example, 
Hobby-Eberly Telescope Dark Energy Experiment (HETDEX) \cite{HETDEX:2008}, 
Dark Energy Spectroscopic Instrument (DESI) \cite{DESI:2013}, 
The Subaru Prime Focus Spectrograph (PFS) \cite{PFS}, 
Wide Field Infrared Survey Telescope (\textit{WFIRST}) \cite{WFIRST:2015}, 
The Large Synoptic Survey Telescope (LSST) \cite{LSST:2012}, 
Spectro-Photometer for the History of the Universe, Epoch of Reionization and Ices Explorer (\textit{SPHEREx}) \cite{Spherex:2014}
and \textit{Euclid} \cite{Euclid:2011}. By locating billions of galaxies, the main goal of these surveys is to provide cosmological probes complementary to the temperature anisotropies and polarization of the cosmic microwave background (CMB). More specifically, combining the two will lead to more precise measurements of cosmological parameters to study, for example, the properties of dark energy, the sum of neutrino masses, and the physics of the early universe.

The increase in the number density of galaxies with these surveys reduces the statistical uncertainties of measuring the galaxy power spectrum and calls for more accurate modeling for extracting more of the cosmological information without modelling systematics.
Traditional LSS analysis focuses on the baryon acoustic oscillations (BAO) feature (e.g. \cite{Bautista:2017, Agathe:2019}), but the constraints on cosmological parameters can be improved with a full shape (FS) analysis which is sensitive to all cosmological parameters \cite{Gil-Marin:2015,Beutler:2016,Ivanov2019}. This allows for galaxy clustering analysis to measure more than just geometrical quantities such as angular diameter distance and the Hubble expansion rate from BAO, but also probe 
the linear growth rate through redshift-space distortions (RSD) \cite{Zhao:2018}, the shape of the primordial power spectrum \cite{Colas:2019}, primordial non-Gaussianity \cite{Zhao:2015}, and neutrino mass \cite{Pellejero-Ibanez:2016}.

To unlock the full potential of the galaxy power spectrum, that is, to use the FS analysis, accurate modeling of the nonlinearities in the galaxy power spectrum is essential. For the high redshift Universe which these surveys are targeting, there are ample quasi-linear regimes in which nonlinear perturbation theory (PT, see \cite{Bernardeau:2002} for a review) accurately models the matter clustering beyond the linear theory \cite{jeong/komatsu:2006,Jeong2009}.  We can, therefore, extend the cosmological analysis using the power spectrum on these quasi-linear scales, and hence improve the cosmological constraints from the surveys. The recent studies \cite{Ivanov2019,Colas:2019} have successfully applied the PT-based analysis for the BOSS DR12 data, and it will only be more powerful with high-redshift galaxy surveys.

Modeling the observed galaxy power spectrum must include two more nonlinearities besides the evolution of the matter density on quasi-linear scales: nonlinear galaxy bias and nonlinear redshift-space distortion. First, what we observe in galaxy surveys is some sampling of the galaxy distribution, which is a biased tracer of the underlying matter density field \cite{Kaiser:1984,BBKS:1986,Fry:1992}. Thanks to the complex gastrophysical nature of the formation and evolution of galaxies, predicting the galaxy distribution from first principles is beyond our reach at present. Instead, on quasi-linear scales, we have created an effective description of galaxy statistics by means of the perturbative bias expansion that includes all possible physical quantities that the galaxy distribution can depend upon. A recent review \cite{Desjacques:review} has presented a complete description of bias by including {\it all} observables that a local observer in the galaxy can measure, at any given order in PT.

In galaxy redshift surveys we infer the distance to the galaxies by their observed spectral shift, assuming that it is solely due to the Hubble flow. The term ``redshift space'' refers to the galaxies position measured in this way. The issue is that peculiar velocities also contribute to the spectral shift and distort the galaxy distribution in redshift space. Because the peculiar velocity is strongly correlated with density field, this effect leads to a systematic change in clustering statistics, so called redshift-space distortions (RSD). Ref.~\cite{Kaiser:87} has established the linear RSD model, and Refs.~ \cite{Heavens:98,Scoccimarro:04,Taruya:2010}, for example, have presented a perturbative description of modeling nonlinear RSD effect.

In addition to the nonlinear galaxy bias and nonlinear RSD, the line-of-sight directional selection effects can further distort the observed galaxy power spectrum. As the name suggests, the selection effect arises because the way that we select the sample of galaxies \citep{Zheng:2011,Hirata:09}. For example, because of the intrinsic alignment \cite{Catelan:2000,Hirata:2004, Brown:2000, Blazek:2011, Joachimi:2011, Martens:2018, Blazek:2015}, galaxies orientation with respect to the line-of-sight direction can be correlated to the large-scale tidal field. If the survey preferentially selects a particular orientation (face-on or edge-on) of galaxies, then the observed galaxy power spectrum can also depend on the line-of-sight directional projection of the large-scale tidal field \cite{Hirata:09}. Another example is for the emission-line selected galaxy samples, where the radiative transfer effects generate a strong-dependence on the line-of-sight directional velocity flow \cite{Zheng:2011}. The bias review \cite{Desjacques:review} has also described a general bias expansion that can be used to take all of the above effects into account at any given order in perturbation theory, resulting in a description of the next-to-leading order (NLO) or 1-loop power spectrum and the leading order (LO) or tree-level bispectrum in a complete bias expansion in \cite{Desjacques:2018}.

The typical expressions for the observed galaxy power spectrum in NLO involves the integrals over the three-dimensional Fourier space, so called one-loop contribution. The naive implementation of these multi-dimensional integrals, however, does not meet the requirements for the data analysis. For example, for the cosmological analysis, the computation of NLO terms needs to be paired with Markov-Chain-Monte-Carlo (MCMC) analysis pipeline that calls the NLO calculation for each set of cosmological parameters. To get robust constraints on cosmological parameters, we typically need parameter chains as long as a few million realizations; that means we need to compute the NLO power spectra millions of times.

This requirement for the data analysis has motivated the development of fast calculation algorithms. 
For example, the FAST-PT methods developed by \cite{Schmittfull:2016} and \cite{McEwen:2016} pre-calculates the angular parts of the loop integrals analytically, leaving the radial parts in terms of spherical Bessel transformations (SBTs). The SBTs can be computed efficiently by using what is commonly known as FFTLog based methods \citep{Siegman:77, TALMAN1978, Hamilton:1999, Gebhardt:2017}. This analysis was extended to higher-order (two-loop) corrections to the matter power spectrum in \cite{FFTPT:2016} and \cite{Slepian:2018}, and to more complicated integration kernels in \cite{Fang:2016}. Recently \cite{Ivanov2019} used a similar technique but for a biased tracer model similar to the one in this work but without the selection effects. \cite{Simonovic:2018} took a different approach by parametrizing the power spectrum as a sum of power laws and factoring out the cosmological dependence so all integrals could be done only once for all sets of parameters. We follow the same procedure as Refs.~\cite{Schmittfull:2016,McEwen:2016} and arrive at a model for the power spectrum of galaxies solely in terms of SBTs, leading to a reduction of computation time per cosmological model by a factor of a thousand (from $\sim$10 minutes to $\sim$1 second) compared to the naive three-dimensional integration using quadrature methods. 

The paper is organized as follows. We start with briefly restating the relevant parts of \cite{Desjacques:2018} in \refsec{bias}, followed by restating a fast method of calculating spherical Bessel transformations in \refsec{fftlog}. Then in \refsec{radial} we describe the transformation of the integrals needed to calculate the NLO power spectrum into the form of a spherical Bessel transformation and give a complete formulation of the redshift-space galaxy power spectrum with FFTLog. In \refsec{tests} we describe various tests of our code to ensure its validity, then in \refsec{fisher} we show the response of the NLO power spectrum to changes in the various bias parameters. 
We conclude in \refsec{conclude}. Following that, \refapp{MO} lists some important coefficient matrices needed to calculate the NLO power spectrum, and in \refapp{newint} we derive some of the fast integral expressions used throughout the work. Lastly in \refapp{math} we give some mathematical identities used throughout the work, and \refapp{qbest} describes our empiric corrections for selecting the optimal biasing parameter.

Throughout this work we use the following conventions and shorthand notations
\be
f(x) \equiv \int \frac{{\rm d}^3\bm k}{(2\pi)^3} f(k) e^{i\bm k \cdot \bm x} \equiv \int_{\bm k}f(k) e^{i\bm k \cdot \bm x}\,,
\ee
\be
\mu_{\bm k, \bm q} = \khat \cdot \qhat\,.
\ee

\section{Formalism: redshift-space galaxy power spectrum including selection effects}\label{sec:bias}
\subsection{General Bias Expansion and The Galaxy Density Contrast in Redshift-Space}
The expression for the 1-loop galaxy power spectrum in redshift space requires the perturbative bias expansion \cite{Desjacques:review} up to third order. Here, we summarize the third-order expression for the observed (redshift-space) galaxy density contrast as derived in \cite{Desjacques:2018}, including the line-of-sight directional selection effects caused by radiative-transfer effects \cite{Zheng:2011} or tidal alignment \cite{Hirata:09}.

Throughout, we work in comoving coordinates ${\bm x}$ and the conformal time variable $\tau$. We denote the matter density contrast $\delta(\bm{x},\tau)$, galaxy density contrast (in real space) $\delta_g (\bm{x},\tau)$, and matter velocity ${\bm v}({\bm x},\tau)$. We also define the scaled matter velocity ${\bm u}({\bm x},\tau) =\frac{1}{{\cal H}(\tau)}{\bm v}({\bm x},\tau)$, where $\mathcal{H} = aH$ is the Hubble expansion rate. We denote the unit vector along the line of sight direction as $\nhat$, and the line-of-sight directional derivative $\partial_\parallel \equiv \nhat^i\partial_i$. This gives the parallel derivative of the scaled line-of-sight velocity $\eta({\bm x},\tau) = \partial_\parallel u_\parallel({\bm x},\tau)$. We denote the matter power spectrum $P^{\delta\delta}(k,\tau)$ while we denote the galaxy power spectrum as $P^{gg}(k,\tau)$ in real space and $P^{gg,s}(k,\mu,\tau)$ in redshift space with 
the line-of-sight directional cosine
$\mu = \khat \cdot \nhat$. 

The key for the general perturbative bias expansion \cite{Desjacques:review} is to expand the galaxy density contrast $\delta_g({\bm x},\tau)$ in 
the following form
\be
\delta_g({\bm x},\tau) = \sum_{\mathcal{O}}[b_{\mathcal{O}}(\tau) + \epsilon_{\mathcal{O}}({\bm x},\tau)]\mathcal{O}(\bm{x},\tau) + \epsilon(\bm {x},\tau),
\label{eq:bias}
\ee
where $\mathcal{O}$ stands for any operator that contributes to the formation and evolution of the galaxies, $b_{\mathcal{O}}$ is the bias parameter associated with that operator. Both $\epsilon_{\cal O}$ and $\epsilon$ stand for stochastic parameters encoding the stochastic processes on {\it sub-grid} scales that are uncorrelated with the operators $\mathcal{O}({\bm x},\tau)$ defined on the large scales where PT is valid. Note that, although only operators at equal time explicitly appear in \refeq{bias}, the expression also includes the effects of all operators along the galaxies' world line, that is, operators at all past times. This is because we can trace the time evolution of operators at each order on large scales where the PT-based models operate.

The central idea behind the perturbative bias expansion in \refeq{bias} is to include all local observables. Following Ref.~\cite{Desjacques:review}, we construct the local gravitational observables in PT starting from the quantity combining the matter density contrast $\delta$ and the tidal field $K_{ij}$ as
\be
\Pi^{[1]}_{ij} = K_{ij} + \frac{1}{3}\delta_{ij}\delta
=
\partial_i\partial_j \Phi \,, 
\ee
where $\Phi$ is proportional to the gravitational potential $\phi$: $\phi({\bm x},\tau)=4\pi Ga^2(\tau)\bar{\rho}_m(\tau)\Phi({\bm x},\tau)$. The superscript $[1]$ here means that the leading order term in $\Pi^{[1]}$ is linear order in PT. Ref.~\cite{Desjacques:review} have demonstrated that one can define the higher-order quantities
\be
\Pi^{[n]}_{ij} = \frac{1}{(n-1)!}\[(\mathcal{H}f)^{-1} \frac{D}{D\tau}\Pi^{[n-1]}_{ij} - (n-1)\Pi^{[n-1]}_{ij} \]\,,
\ee
which capture all local gravitational observables. Here, $D/D\tau = \partial_\tau + v^i\partial_{x,i}$ is the convective derivative following the peculiar velocity field. 
The $n$-th order rank-2 tensors $\Pi^{[n]}_{ij}$, therefore, form our building blocks for the perturbative bias expansion. Taking every combination up to 3rd order, we find the following set of operators 
\be
\left\{
\delta, ~\delta^2,~ \delta^3,~ \text{tr}[KK] = K^2,~ \delta K^2,~ K^3, O_{\rm td} 
\right\}\,,
\label{eq:bias_real}
\ee
suffices the description of galaxy clustering to third order, or NLO in galaxy power spectrum. Here,
\be
O_{\rm td} 
= \frac{8}{21}K_{ij}
\left(
\frac{\partial_i\partial_j}{\nabla^2}
-
\frac13 \delta_{ij}
\right)
\left(
\delta^2 - \frac{3}{2}K^2
\right)\,,
\ee
that appears in third order is the lowest order non-trivial quantity, which cannot be formed by algebraic combination of $\delta$ and $K_{ij}$, of galaxy bias expansion. It is, however, clearly a local observable, as $O_{\rm td}$ is proportional to the convective derivative of the tidal field.

In addition to the deterministic bias expansion above, we also include the stochastic contribution to the galaxy power spectrum given as
\be
P_\epsilon(k) = P_\epsilon^{\{0\}} + k^2P_\epsilon^{\{2\}} + \mathcal{O}(k^4),
\ee
and the higher derivative bias terms that incorporate the feedback from the small-scale dynamics, which is simply $b_{\nabla^2 \delta}\nabla^2 \delta$ at third order.

Transferring to redshift space, we need to model the peculiar velocity field of galaxies, which coincides, to linear order, with the matter density field. We also include the velocity bias, deviation of galaxy velocity field from the matter velocity field, with additional higher derivative bias parameters $\beta_{\nabla^2 \bm v}$ and $\beta_{\partial_\parallel^2 \bm v}$, as
\be
\bm v_g = \bm v + \beta_{\nabla^2 \bm v}\nabla^2 \bm v + \beta_{\partial_\parallel^2\bm v}\partial_\parallel^2 \bm v + \varepsilon_v\,.
\ee
In this notation, we write the line-of-sight directional velocity divergence as
\be
\eta_g = \(1-\beta_{\nabla^2 \bm v}k^2 - \beta_{\partial_\parallel^2 \bm v}k^2\mu^2\)\eta + \varepsilon_\eta,
\ee
where $\mu = \khat\cdot\nhat$. The coordinate transformation between the real space and the redshift space is 
\be
{\bm x}_s = {\bm x} + u_\parallel \nhat \,,
\ee
and the number of galaxies stays invariant under the coordinate transformation:
\be
\left(1+\delta_{g,s}({\bm x}_s)\right) d^3x_s
=
\left(1+\delta_{g}({\bm x})\right) d^3x
\,.
\label{eq:RSDdensity}
\ee
Here, we neglect the terms proportional to $1/r$ in favor of $\partial_\parallel$ whose contribution dominates on the quasi-linear scales where NLO terms are important. By expressing \refeq{RSDdensity} at the redshift-space coordinates, we find the expression for the galaxy density contrast to third order
\ba
\label{eq:delta_gs}
\delta_{g,s} &= \delta^{\rm Jac}_g + \delta^{\rm disp}_g \\
\delta^{\rm Jac}_g &= (1+\delta_g)(1-\eta_g + \eta_g^2) - \eta_g^3 - 1 \\
\delta^{\rm disp}_g &= -u_{g\parallel}\partial_\parallel\delta^{\rm Jac}_g + \frac{1}{2}u^2_{g\parallel}\partial^2_{\parallel}\delta^{\rm Jac}_g + (u_{g\parallel}\partial_\parallel u_{g\parallel})\partial_\parallel \delta^{\rm Jac}_g \,.
\ea
Note that the $\delta_g^{\rm Jac}$ terms corresponds to the Jacobian ($\partial {\bm x}/\partial {\bm x}_s$) of the coordinate mapping to redshift-space, and the $\delta_g^{\rm disp}$ terms corresponds to the displacement of the fields from the real space coordinate ${\bm x}$ to the observed redshift coordinate ${\bm x}_s$.

In order to include the line-of-sight dependent selection effects, which treat the line of sight $\nhat$ as a preferred direction, we need to employ additional bias terms constructed by combining the local gravitational observables $\Pi^{[n]}_{ij}$ with the line-of-sight directional unit vector $\nhat$, allowing for combinations such as $\Pi_\parallel = \Pi_{ij} \nhat^i \nhat^j$. To third order in PT, the additional terms are:
\be
\left\{ \eta, ~\delta \eta, ~(KK)_\parallel, ~\eta^2, ~\Pi^{[2]}_\parallel, ~\delta\Pi^{[2]}_\parallel,
~(K\Pi^{[2]})_\parallel, ~\eta\Pi^{[2]}_\parallel, ~\Pi^{[3]}_\parallel, ~\partial_\parallel^2\delta, ~\nabla^2\eta, ~\partial_\parallel^2\eta \right\}\,.
\label{eq:bias_selection}
\ee
Note that we count the last three higher-derivative terms as third order as they are suppressed by a factor of $(k/k_{\rm NL})^2$ compared to the respective linear order quantities \cite{Desjacques:review}.

\subsection{Redshift-Space Galaxy Power Spectrum}
Combining all contributions we have discussed in the previous section, we find the expression for the one-loop (adding LO and NLO) galaxy power spectrum in redshift space as follows. Following the convention in \cite{Desjacques:2018}, we organize the final result in the following way:
\be
P_{\rm LO+NLO}^{gg,s}(k,\mu) = P_{l+hd}^{gg,s}(k,\mu) + P_{22}^{gg,s}(k,\mu) + 2P_{13}^{gg,s}(k,\mu)\,.
\label{eq:Pkgs}
\ee
Here, we absorb all of the non-integral terms into a single term $P_{l+hd}^{gg,s}(k,\mu)$ which contains the leading order (LO) Kaiser terms, the stochastic terms and the higher derivative terms:
\ba
P_{l+hd}^{gg,s}(k,\mu) =& \(b_1 - b_\eta f\mu^2\)^2P_L(k) + P_\epsilon^{\{0\}} 
+ k^2P_\epsilon^{\{2\}} + \mu^2k^2b_\eta P_{\epsilon\varepsilon_\eta}^{\{2\}}
\vs
& - 2\[b_1b_{\nabla^2\delta} - \mu^2fb_\eta\(b_{\nabla^2\delta} + b_1\beta_{\nabla^2\bm v} + b_1\beta_{\partial^2_\parallel \bm v}\mu^2\)
+\mu^4f^2b_\eta^2\(\beta_{\nabla^2 \bm v} + \beta_{\partial_\parallel^2 \bm v}\mu^2\)\]k^2P_L(k)\,.
\label{eq:non-integral}
\ea

We further divide the rest of the NLO terms as $P_{22}^{gg,s}(k,\mu)$ that comes from the multiplication of two second order quantities, and $P_{13}^{gg,s}(k,\mu)$ that comes from the multiplication of linear order quantities and third order quantities. Including all local and selection observables in \refeq{bias_real} and \refeq{bias_selection}, there are 16 deterministic bias parameters to begin with.

The expression for $P_{22}^{gg,s}(k,\mu)$ may be written as
\be
P_{22}^{gg,s}(k,\mu) = \sum\limits_{{\cal O},{\cal O}'\in\mathfrak{D}_{2}} b_{\cal O}b_{{\cal O}'}\mathcal{I}^{{\cal O},{\cal O}'}(k,\mu)\,,
\ee
where the summation runs over all second order terms $\mathfrak{D}_{2}$ contributing to $\delta_{g,s}$ in \refeq{delta_gs}:
\be
\mathfrak{D}_{2} = \left\{\delta^{(2)},\eta^{(2)},\delta^2,K^2,\delta\eta,\eta^2, (KK)_\parallel, \Pi^{[2]}_\parallel,u_\parallel\partial_\parallel\delta,u_\parallel\partial_\parallel\eta \right\}\,,
\label{eq:D2}
\ee
with associated coefficients for the second order contributions that we call $b_{\cal O}$
\be
\{b_{\cal O}\}_{\mathfrak{D}_{2}} = 
\left\{ b_1, b_\eta, b_{\delta^2} = b_2/2, b_{K^2}, b_{\delta\eta}, b_{\eta^2}, b_{(KK)_\parallel}, b_{\Pi^{[2]}_\parallel}, -b_1, -b_\eta \right\}\,.
\ee
The functions $\mathcal{I}^{{\cal O},{\cal O}'}$ are the two-point correlators of the second-order operators:
\be
\left<
{\cal O}({\bm k}){\cal O}'({\bm k})
\right>
=
(2\pi)^3
\mathcal{I}^{{\cal O},{\cal O}'}(k,\mu)
\delta^{D}({\bm k}+{\bm k}')\,,
\ee
and one can find the explicit formula of ${\cal I}^{{\cal O},{\cal O}'}(k,\mu)$ in terms of a loop integration over two linear power spectra in Ref.~\cite{Desjacques:2018}. Taking all binary combinations of 10 terms in the second order expansion (\refeq{D2}), one might expect that we need to compute 55 different ${\cal I}^{{\cal O},{\cal O}'}(k,\mu)$ terms. Ref.~\cite{Desjacques:2018}, however, further reduces the number of integrals, ends up finding that
\be\label{eq:P22ggs_old}
P_{22}^{gg,s}(k,\mu)
=
\sum_{n=0}^4\sum_{(m,p)}
{\cal A}_{n(m,p)}
(f,\left\{b_{\cal O}\right\}_{\mathfrak{D}_2})
{\cal I}_{mp}(k)\mu^{2n}\,,
\ee
with 
\be
{\cal I}_{mp}(k)
\equiv
2 \left[
\int_{\bm q}
\frac{q^{p-2}k^{6-p}}{|{\bm k}-{\bm q}|^4}
\mu_{\khat,\qhat}^m
P_{\rm L}(q)
P_{\rm L}(|{\bm k}-{\bm q}|)
-
\frac{\delta_{p6}}{m+1}
\left(
\int_{\bm q} \left[P_{\rm L}(q)\right]^2
\right)
\right]\,.
\ee
It turns out that 23 combinations of $(m,p)$ pairs suffices for the calculation of $P_{22}^{gg,s}(k,\mu)$. 

Note that this form of $P_{22}^{gg,s}(k,\mu)$ does not suit the FFTLog-based fast calculation method that we are developing in this paper, and we develop an alternative expression using the Hankel transformation in \refsec{radial}.

The expression for $P_{13}^{gg,s}(k,\mu)$ takes a similar form
\be \label{eq:p13ggs}
P_{13}^{gg,s}(k,\mu) = \sum\limits_{{\cal O}\in\mathfrak{D}_{3}}\(b_1 - b_\eta f\mu^2\)b_{\cal O}f^{n_f({\cal O})}f_{\cal O}(k,\mu)P_L(k)\,.
\ee
The NLO term $P_{13}^{gg,s}(k,\mu)$ is constructed from multiplying third order quantities with the linear order quantities that are encoded in the $(b_1 - b_\eta f\mu^2)$ term in the expression. The summation runs over the third order contributions $\mathfrak{D}_3$, which are
\ba
\mathfrak{D}_{3} = &\left\{\delta^{(3)}, \eta^{(3)},2\text{tr}[KK^{(2)}], \delta\eta^{(2)}, 2\eta\eta^{(2)}, 2(KK^{(2)})_\parallel, O_{td}, \delta\Pi^{[2]}_\parallel, \eta\Pi^{[2]}_\parallel, (\Pi^{[2]}K)_\parallel, s^k\partial_k\Pi^{[2]}_\parallel, \right. 
\vs
 & \left. u_\parallel^{(2)}\partial_\parallel\delta, u_\parallel^{(2)}\partial_\parallel\eta, u_\parallel\partial_\parallel\eta^{(2)}, u_\parallel\partial_\parallel\Pi^{[2]}_\parallel, \Pi^{[3]}_\parallel \right\}\,,
\ea
with corresponding coefficients $\{b_{\cal O}\}_{\mathfrak{D}_{3}}$ in the third order expressoin of $\delta_{g,s}$:
\ba
\{b_{\cal O}\}_{\mathfrak{D}_{3}} 
= &\left\{b_1, b_\eta, b_{K^2}, b_{\delta\eta}, b_{\eta^2},\vphantom{b_{\Pi^{[3]}_\parallel}} b_{(KK)_\parallel}, b_{td}, b_{\delta\Pi^{[2]}_\parallel}, b_{\eta\Pi^{[2]}_\parallel}, b_{(\Pi^{[2]}K)_\parallel}, -b_{\Pi^{[2]}_\parallel}, \right.
 \vs
 &\left.-b_1, -b_\eta, -b_\eta, -b_{\Pi^{[2]}_\parallel}, b_{\Pi^{[3]}_\parallel} + 2b_{\Pi^{[2]}_\parallel} \right\}\,.
\ea
Note that the set $\mathfrak{D}_3$ excludes the third order contributions coming from the product of three first-order operators. This is because we absorb their contribution into the coefficients of \refeq{non-integral} by renormalization (see 
App. C.1 of Ref.~\cite{Desjacques:2018} for the details).
In addition to the bias parameters, each velocity-oriented operator in $\mathfrak{D}_3$ is multiplied with the linear growth rate $f$ with the power denoted as $n_f({\cal O})$ in \refeq{p13ggs}. This power is the same as the number of velocity terms (either $\eta$ or $u_\parallel$) in the operator:
\be
\{n_f({\cal O})\}_{\mathfrak{D}_{3}} = \{0,1,0,1,2,0,0,0,1,0,0,1,2,2,1,0\}\,,
\ee
in the same order as the previous two sets.

Finally, the function 
\be
f_{\cal O}(k,\mu) = (1, \mu^2, \mu^4) \bm{\mathcal{M}}({\cal O}) 
\begin{bmatrix}
\mathcal{I}_1(k) \\
\mathcal{I}_2(k) \\
\mathcal{I}_3(k) \\
\mathcal{I}_4(k) \\
\mathcal{I}_5(k)
\end{bmatrix}\,,
\ee
contains the loop integrals $\mathcal{I}_{n}(k)$, which are (see App. D of \cite{Desjacques:2018} for the details)
\ba \label{eq:Ii}
\mathcal{I}_1(k) 
&= 
k^2 \int_{\bm q} 
\frac{\bm k \cdot \bm q}{q^2 |\bm k - \bm q|^2} 
(1 - \mu_{\bm k,\bm q}^2) 
P_L(q) 
\vs
\mathcal{I}_2(k) 
&= 
k^2 \int_{\bm q} 
\frac{q^2}{q^2 |\bm k - \bm q|^2} 
(1 - \mu_{\bm k,\bm q}^2) 
P_L(q)
\vs
\mathcal{I}_3(k) 
&= 
\int_{\bm q} 
\[\frac{q^2}{|\bm k - \bm q|^2} 
(1-\mu_{\bm k,\bm q}^2) 
- \frac{2}{3} \]
P_L(q)
\vs
\mathcal{I}_4(k) 
&= 
\int_{\bm q} 
\[\frac{(\khat \cdot \bm q )^2}{|\bm k - \bm q|^2}
(1-\mu_{\bm k,\bm q}^2) 
- \frac{2}{15} \]
P_L(q)
\vs
\mathcal{I}_5(k) 
&= 
\int_{\bm q} 
\[\frac{(\khat \cdot \bm q )^4}{q^2|\bm k - \bm q|^2}
(1-\mu_{\bm k,\bm q}^2) 
- \frac{2}{35} \]
P_L(q)\,.
\ea
Here, $\bm{\mathcal{M}}({\cal O})$ is the $3 \times 5$-coefficient matrix defined for each third-order operator. We present the matrices in \refapp{MO}.

The main equation \refeq{Pkgs} combined with each component in \refeq{non-integral}, \refeq{P22ggs_old}, and \refeq{p13ggs} completes the expression for observed (redshift-space) galaxy power spectrum to one-loop (LO+NLO) order. To compute the one-loop power spectrum, we need to evaluate the 28 (23 for $I_{mp}(k,\mu)$ and 5 for $I_n(k)$) two-dimensional integrals. The remaining task of this paper is to reduce the computational burden by reducing them into the one-dimensional integrals which can be carried out faster by using the FFTlog-based method
\citep{Siegman:77, TALMAN1978, Hamilton:1999, Gebhardt:2017}.

\section{FFTLog Transformation}\label{sec:fftlog}
The FFTLog-based integration accelerates the computation speed of the spherical Bessel transformations, SBTs, sometimes known as Hankel transformations. For the implementation of the one-loop power spectrum expression, we only need the SBTs involving a single Bessel function:
\be
\xi^\ell_n(r) = \int_0^\infty
\frac{k^2 {\rm d}k}{2\pi^2} k^n j_\ell(kr)P_L(k).
\label{eq:xiLn}
\ee
Due to the oscillatory nature of the integrand, mainly caused by the spherical Bessel functions, these integrals are often slow to compute with ordinary quadrature methods. 

The key observation \citep{Siegman:77, TALMAN1978, Hamilton:1999, Gebhardt:2017} for the fast integration of \refeq{xiLn} is to perform the integration in the logarithmic space. That is, defining 
\be
k = k_0e^\kappa \qquad r=r_0e^\rho\,,
\ee
reduces \refeq{xiLn} to a convolution integral in $\kappa$ and $\rho$:
\ba
\xi^\ell_n(r) = &\frac{k_0^3e^{-\rho q}(k_0r_0)^n}{2\pi^2}\int_{-\infty}^\infty {\rm d}\kappa\, e^{\kappa(3-q+n)}
P_L(k_0e^\kappa) e^{q(\kappa+\rho)}j_\ell(k_0r_0e^{\kappa+\rho})\,,
\ea
which can be instead performed as a multiplication in the Fourier-dual space of $\kappa$. Here, we introduce a power law biasing, $(kr)^q$, to enhance the performance of the numerical implementation, more specifically, to reduce the aliasing effect.
Explicitly, we define the one-dimensional Fourier transform of the biased power spectrum and the biased spherical Bessel function, respectively, as
\ba \label{eq:phidef}
\phi^q(t) =& \int_{-\infty}^\infty \frac{{\rm d}\kappa}{2\pi}e^{i\kappa t}e^{\kappa(3-q)}P(k_0e^\kappa)\,,
\\ \label{eq:Mlqdef}
e^{q(\kappa + \rho)}j_{\ell}(k_0r_0e^{\kappa + \rho}) =& \int\limits_{-\infty}^\infty \frac{{\rm d}t}{2\pi} e^{i(\kappa + \rho)t}M_{\ell}^q(t)\,,
\ea
where inverting \refeq{Mlqdef} serves as the definition of $M_\ell^q$. We use \refeqs{phidef}{Mlqdef} to  re-write \refeq{xiLn} as the integration in the dual ($t$) space as
\be
\xi_n^\ell(r) = \frac{k_0^3e^{-\rho q}(k_0r_0)^n}{2\pi^2}\int\limits_{-\infty}^\infty \frac{{\rm d}t}{2\pi}e^{i\rho t}\phi^{q-n}(t)M_\ell^q(t) \,.
\ee

The Fourier transform of the biased spherical Bessel function, $M_\ell^q(t)$, can be defined analytically in terms of Gamma functions:
\be
M_\ell^q(t) = 2^{n-1} \sqrt{\pi} (k_0r_0)^{it - q} \frac{\Gamma[\frac{1}{2}(\ell+q-it)]}{\Gamma[\frac{1}{2}(3+\ell-q+it)]}\,.
\ee
So calculating $\xi^\ell_n(r)$ amounts to just another Fourier transform of $\phi^{q-n}(t)M_\ell^q(t)$ using FFT.

In order to implement the FFTlog-based method, we need to set three parameters: $k_0$, $r_0$ and $q$. Following the discussion in \cite{Hamilton:1999}, we set $k_0$ and $r_0$ so that $k_0r_0\approx 1$. The choice of the biasing parameter $q$ is more subtle. While Ref.~\cite{Gebhardt:2017} have systematically studied the choice of the biasing parameter, their prescription of {\it choosing a $q$ value to make the slopes at the end of the input equal} only applies for calculating transformations of the linear power spectrum. On the other hand, calculating the one-loop power spectrum that we consider here requires the FFTlog transform of various other types of functions. We therefore extend the prescription of Ref.~\cite{Gebhardt:2017}, primarily by introducing empirical corrections based on the input function. We present the extended prescription in \refapp{qbest}. 
Besides the choice of the biasing parameter, all of our FFTLog computations use the implementation of \cite{Gebhardt:2017}.

\section{Redshift-Space Galaxy Power Spectrum with FFTLog}\label{sec:radial}
In this section, we present the details of our implementation
of the redshift space one-loop power spectrum using the 
FFTlog transformation in \refsec{fftlog}.

\subsection{\texorpdfstring{$P_{22}(k)$}{P22}}\label{sec:P22_fftlog}
Although the expression \refeq{P22ggs_old} is compact with only 23 ${\cal I}_{mp}(k)$ integrals, we find it difficult to manipulate ${\cal I}_{mp}(k)$ integrals into a form suitable for the FFTLog transformation.
To take advantage of the FFTLog transformation, instead, we start from the second order kernel for the redshift-space density contrast as presented in Eq.~(86) of \cite{Desjacques:2018}:
\ba
Z_2&(\bm q_1, \bm q_2) = 
\frac{1}{2} b_2 + \frac{1}{9}b_{(KK)_\parallel} - \frac{1}{3} b_{K^2} + \frac{5}{7} \(b_1 + b_{\Pi_\parallel^{[2]}}\mu^2 \) - \frac{3}{7}fb_\eta\mu^2
\vs
&+ \frac{1}{2} \( b_1 - fb_\eta \mu^2\)\frac{k^2 \bm q_1\cdot \bm q_2}{q_1^2q_2^2}
+ \[b_{K^2} - \frac{5}{7} b_1  + \(\frac{3}{7}fb_\eta - \frac{5}{7}b_{\Pi_\parallel^{[2]}} \)\mu^2 \]\frac{(\bm q_1 \cdot \bm q_2)^2}{q_1^2q_2^2}
\vs
&+ \(b_{\Pi_\parallel^{[2]}} + b_{(KK)_\parallel}\) \frac{(\bm q_1 \cdot \bm q_2)q_{1z}q_{2z}}{q_1^2q_2^2} - \frac{1}{6}\(3f(b_{\delta\eta} + b_1) + 2b_{(KK)_\parallel} \)\frac{q_{1z}^2q_2^2 + q_1^2q_{2z}^2}{q_1^2q_2^2} + f^2(b_{\eta^2} + b_\eta)\frac{q_{1z}^2q_{2z}^2}{q_1^2q_2^2}
\vs
&+ \frac{(fk\mu)^2}{2}\frac{q_{1z}q_{2z}}{q_1^2q_2^2} + \frac{fk\mu}{2} \[\frac{q_{1z}}{q_1^2} \(b_1 - f(b_\eta + 1) \frac{q_{2z}^2}{q_2^2} \)  + \frac{q_{2z}}{q_2^2} \(b_1 - f(b_\eta + 1) \frac{q_{1z}^2}{q_1^2} \) 
\]\,,
\ea
where we define $q_{iz} = q_i \mu_{\bm n, \bm q_i} 
= q_i (\nhat \cdot \qhat_i)$.
With the kernel $Z_2$, the expression for $P_{22}^{gg,s}(k,\mu)$ becomes
\ba
P_{22}^{gg,s}(k,\mu) 
=& 
2\int_{\bm q} [Z_2(\bm q, \bm k - \bm q)]^2 P_L(q) P_L(|\bm k - \bm q|)
-
2\int_{\bm q} [Z_2(\bm q, - \bm q)P_L(q) ]^2 
\vs
=& 2 (2\pi)^3
\int_{\bm p} 
\int_{\bm q} 
[Z_2(\bm p, \bm q)]^2 P_L(p)P_L(q)
\delta^D(\bm p + \bm q - \bm k)
-
2\int_{\bm q} [Z_2(\bm q, - \bm q)P_L(q) ]^2\,.
\ea
Note that we subtract the constant term that renormalizes the shot-noise contribution $P_0$.
Next, we expand $[Z_2\(\bm p,\bm q\)]^2$, separating the angular dependence in terms of Legendre polynomials for each of the angles in the kernel, $\nhat\cdot\qhat$, $\nhat\cdot\phat$, and $\phat\cdot\qhat$. The expression for $P_{22}^{gg,s}(k)$ then becomes the linear combination
\be \label{eq:p22intform}
P_{22}^{gg,s}(k,\mu) = 2\sum_{abcn_1n_2} \mathcal{C}_{abc}^{n_1n_2}(k,\mu,f, b_O) \mathcal{I}^{n_1n_2}_{abc}(k,\mu)\,,
\ee
with the coefficients ${\cal C}_{abc}^{n_1n_2}$ and the integral
\be
\mathcal{I}_{abc}^{n_1n_2}(k,\mu) = (2\pi)^3\int_{\bm q}\int_{\bm p} q^{n_1-2}p^{n_2-2}\delta_D(\bm p+\bm q-\bm k)P_L(q)P_L(p)\mathcal{L}_{a}\(\nhat\cdot\qhat \) \mathcal{L}_{b}\(\nhat\cdot \phat\) \mathcal{L}_{c}\(\phat\cdot\qhat \)\,.
\ee
The angular integral can be further simplified to yield
\ba 
\label{eq:mediumint}
\mathcal{I}^{n_1n_2}_{abc}(k,\mu) 
=& (2\pi)^3(-1)^{a+b+c}\sum\limits_{\ell_r} \mathcal{L}_{\ell_r}(\mu)(2\ell_r+1) \tj{a}{b}{\ell_r}{0}{0}{0} 
\sum\limits_{\ell_a\ell_b} i^{\ell_a+\ell_b-\ell_r}(2\ell_a+1)(2\ell_b+1)
\vs
&\times \tj{a}{\ell_a}{c}{0}{0}{0} \tj{b}{\ell_b}{c}{0}{0}{0} \tj{\ell_r}{\ell_a}{\ell_b}{0}{0}{0} \sj{a}{b}{\ell_r}{\ell_b}{\ell_a}{c} 
\int \frac{dr}{2\pi^2}r^2j_{\ell_r}(kr) 
\xi^{\ell_a}_{n_1-2}(r)\xi^{\ell_b}_{n_2-2}(r)\, ,
\ea
with a Wigner-3j symbol
$\tj{a}{b}{c}{0}{0}{0}$,
a Wigner-6j symbol
$\sj{a}{b}{c}{d}{e}{f}$, and $\xi_n^\ell(r)$ defined in \refeq{xiLn}. We present the detailed derivation of the angular integration in \refapp{newint}.

Note that the Wigner symbols in \refeq{mediumint} dictates that (A) $\ell_a + \ell_b - \ell_r$ is even which guarantees the integrand is real, and (B) the values of $\ell_a$, $\ell_b$, and $\ell_r$ are bounded by triangle conditions, for example, $|\ell_r - \ell_a| \leq \ell_b \leq \ell_r + \ell_a$, for any permutation and likewise for every other 3j symbol. We refer the readers to Ref.~\cite{textbook} for the other properties of Wigner symbols.

The coefficients 
${\cal C}_{abc}^{n_1n_2}$ 
are too lengthy to list in the paper, and we present them in the supplementary material \cite{Supplementary}. In total there are 51 unique coefficients for the 83 different possible combinations of indices when taking into account the symmetry of the expression, $\mathcal{I}^{n_1n_2}_{abc} = \mathcal{I}^{n_2n_1}_{bac}$. Implementation of \refeq{p22intform} along with \refeq{mediumint} requires 98 FFTlog computations.

To reduce the number of FFTlog transformations, we manipulate \refeq{p22intform} and \refeq{mediumint} such that the FFTlog ($r$-integration) operation takes place only at the last step. That is, for a given combination of $k^n{\cal L}_\ell(\mu)j_{\ell_r}(kr)$, we pre-compute all {\it internal} summations in \refeq{p22intform} and \refeq{mediumint} so that the final expression for the $P_{22}^{gg,s}(k,\mu)$ becomes 
\be
P_{22}^{gg,s}(k,\mu) = 
2(2\pi)^3
\sum\limits_{\ell=0}^4
\mathcal{L}_{2\ell}(\mu)
\sum\limits_{n=0}^4 
k^n 
\sum\limits_{\ell_r=0}^8
\int\frac{{\rm d}r}{2\pi^2}r^2
j_{\ell_r}(kr) 
\mathcal{M}_{\ell_r}^{n,\ell}(f, b_{\cal O},r)\,.
\label{eq:finalP22}
\ee
Here, $\mathcal{M}$ contains summation over Wigner symbols and $\xi_n^\ell(r)$ functions and depends on the parameters such as $f$ and $b_{\cal O}$. We have also absorbed the renormalization contributions into ${\cal M}_{\ell_r}^{0,\ell}$\,. Again, the expression for ${\cal M}$ is very lengthy, so we present them only in the supplementary material \cite{Supplementary}. The final expression in \refeq{finalP22} reduces the number of FFTlog transformation down to 73, a significant decrease from the earlier method using \refeq{mediumint}. For the numeric calculations in this work, and the code we provide, therefore, we use this form of $P_{22}^{gg,s}(k,\mu)$.

\subsection{\texorpdfstring{$P_{13}(k)$}{P13}}\label{sec:P13_fftlog}

To transform the 1-3 integrals in \refeq{p13ggs} into the numerically faster form of \refeq{xiLn} we first factor them into radial and angular components, then do the angular integral analytically. This leaves us with just a radial integral which is in the form of a spherical Bessel transformation and can be done very quickly with FFTLog. 
That is, we can directly transform these integrals, $\mathcal{I}_i(k)$ defined in \refeq{Ii}, with an identity from Ref.~\cite{Schmittfull:2016},

\be\label{eq:svm13}
\int_{\bm q} \frac{1}{|\bm k - \bm q|^2}q^n (\khat\cdot\qhat)^\ell P_L(q)
=
\sum\limits_{\ell' = 0}^\ell (2\ell' + 1) \alpha_{\ell\ell'}
\int_0^\infty {\rm d}r\, r j_{\ell'}(kr)\xi^{\ell'}_n(r)
\equiv 
\sum\limits_{\ell' = 0}^\ell (2\ell' + 1) \alpha_{\ell\ell'}
\mathcal{P}_{13}^{\ell,n}(k) 
\,,
\ee
with
\begin{align}
\alpha_{\ell\ell'} = 
\begin{cases}
\frac{\ell!}
{2^{(\ell-\ell')/2}  [(\ell-\ell')/2]! (\ell+\ell'+1)!!} & \text{if $\ell \geq \ell'$ and $\ell$ and $\ell'$ are both even or odd.}  \\
0 & \text{otherwise.}
\end{cases}
\end{align}
Applying this identity to the five integrals in \refeq{Ii} results in
\ba \label{eq:P13rad}
\mathcal{I}_1(k) 
&= 
\frac{2k^3}{5}\(\mathcal{P}_{13}^{1,-1}(k) - \mathcal{P}_{13}^{3,-1}(k)\)
\vs
\mathcal{I}_2(k) 
&= 
\frac{2k^2}{3}\(\mathcal{P}_{13}^{0,0}(k) - \mathcal{P}_{13}^{2,0}(k)\)
\vs
\mathcal{I}_3(k) 
&= 
\frac{2}{3}\(\mathcal{P}_{13}^{0,2}(k) - \mathcal{P}_{13}^{2,2}(k)\) 
\vs
\mathcal{I}_4(k) 
&=
\frac{2}{15}\mathcal{P}_{13}^{0,2}(k) + \frac{2}{21}\mathcal{P}_{13}^{2,2}(k) - \frac{8}{35}\mathcal{P}_{13}^{4,2}(k)
\vs
\mathcal{I}_5(k) 
&=
\frac{2}{35}\mathcal{P}_{13}^{0,2}(k) + \frac{2}{21}\mathcal{P}_{13}^{2,2}(k) - \frac{32}{385}\mathcal{P}_{13}^{4,2}(k) - \frac{16}{231}\mathcal{P}_{13}^{6,2}(k)\,.
\ea
Using these identities, we can calculate all of the integrals required for $P_{13}^{gg,s}(k,\mu)$ with 16 unique FFTLog transformations. 
It is worth noting that we have dropped the renormalization terms present in the original integrals, for example, in $I_3(k)$, $I_4(k)$ and $I_5(k)$. This is because FFTLog is immune to the constant ($k$-independent) contributions which requires the inclusion of $q=0$ ($\log q=-\infty$).

The expressions for $P_{22}^{gg,s}$ and $P_{13}^{gg,s}$ have exactly four overlapping FFTLog transformations, resulting in a final total of 85 for the entire one-loop power spectrum model in the general bias expansion. Despite the seemingly large number of integrals that need to be done, this method is about a factor of thousand faster than using those integrals in Ref.~\cite{Desjacques:2018}, for example, going from $\sim$10 minutes to $\sim$1 second per power spectrum model on a 3.2 GHz Intel CPU with our {\sf Julia} implementation.  

\subsection{Multipole Expansion} \label{sec:multipole}
We decompose the line-of-sight angle dependence of the redshift-space power spectrum by expanding the $\mu$-dependence into Legendre polynomials. When considering statistically homogeneous density and velocity fields at a constant time, the forward-directional velocity field is statistically in-distinguishable from the backward-directional velocity field; hence, the redshift-space power spectrum in this case only contains even power in $\mu$. This case must be contrasted with the real universe where large-scale structure evolves along the line-of-sight direction, and such evolution generates odd-multipoles in the redshift-space power spectrum. This effect, however, is suppressed by a factor of $1/kr$ where $r$ is the distance to the galaxy survey volume \cite{Kaiser:87}.

We denote the even-order Legendre multipoles as 
\ba
&P_{\rm LO+NLO}^{gg,s}(k,\mu)
= \sum\limits_{\ell=0}^4 P_{{\rm LO+NLO},2\ell}^{gg,s}(k) \mathcal{L}_{2\ell}(\mu) 
\equiv  
\sum\limits_{\ell=0}^4  \[P_{l+hd,2\ell}^{gg,s}(k) + P_{22,2\ell}^{gg,s}(k) + 2P_{13,2\ell}^{gg,s}(k)\]\mathcal{L}_{2\ell}(\mu)\,,
\ea
where 
\be
P_{X,\ell}^{gg,s}(k) = \frac{2\ell + 1}{2}\int_{-1}^1 {\rm d}\mu\, \mathcal{L}_\ell(\mu)P_X^{gg,s}(k,\mu).
\ee
We find the Legendre multipoles for the linear and higher derivative terms in \refeq{non-integral} as
\ba \label{eq:lhdpole}
P_{l+hd,0}^{gg,s}(k) &= b_1^2P_L(k) - \frac{2}{15}b_1P_L(k)\[fb_\eta(-5\beta_{\nabla^2\bm v}k^2 - 3\beta_{\partial_\parallel^2 \bm v}k^2 + 5) + 15b_{\nabla^2\delta}k^2\] 
\vs
&- \frac{1}{35}f^2P_L(k)b_\eta^2\[14\beta_{\nabla^2 \bm v}k^2 +10\beta_{\partial_\parallel^2 \bm v}k^2 - 7\] + \frac{1}{3}k^2b_\eta\[2b_{\nabla^2\delta}fP_L(k) + P^{\{2 \}}_{\epsilon\varepsilon_\eta}\]
\vs
& + k^2P^{\{2\}}_\epsilon + P^{\{0\}}_\epsilon
\vs
P_{l+hd,2}^{gg,s}(k) &= \frac{2}{21}b_\eta  \left\{ 2fP_L(k)\[fb_\eta\(3-k^2(6\beta_{\nabla^2\bm v} + 5 \beta_{\partial^2_\parallel \bm v})\) + b_1\(k^2(7\beta_{\nabla^2\bm v} + 6\beta_{\partial^2_\parallel\bm v}) - 7\)\] \right.
\vs
& \left. + 7k^2(2b_{\nabla^2\delta}fP_L(k) + P^{\{ 2\}}_{\epsilon\varepsilon_\eta}) \right\}
\vs
P_{l+hd,4}^{gg,s}(k) &= -\frac{8}{385}fP_L(k)b_\eta \[ fb_\eta \( 22\beta_{\nabla^2\bm v}k^2 + 30\beta_{\partial^2_\parallel\bm v}k^2 - 11 \) - 22 b_1\beta_{\partial^2_\parallel\bm v}k^2\]
\vs
P_{l+hd,6}^{gg,s}(k) &= -\frac{32}{231}\beta_{\partial^2_\parallel\bm v}f^2k^2P_L(k)b^2_\eta.
\ea
For the 1-3 loop terms we get
\be
P_{13,\ell}^{gg,s}(k) = \sum\limits_{n=1}^5 \mathcal{C}_n^{1-3,\ell}(f,\{b_{\cal O}\}_{\mathfrak{D}_3})\mathcal{I}_n(k)P_L(k),
\ee
where $\mathcal{C}_n^{1-3,\ell}$ is a coefficient matrix listed in the supplementary material of \cite{Desjacques:2018}. The 2-2 loop terms are already in the proper format for multipole decomposition in \refeq{finalP22}.

\section{Numerical implementation}\label{sec:tests}
For the non-linear redshift-space power spectrum, we have implemented \refeq{non-integral}, \refeq{finalP22}, and \refeq{p13ggs} along with \refeq{P13rad} in {\sf Julia}. For the FFTlog transformations, we use the implementation of the TwoFAST module \cite{Gebhardt:2017}. Our {\sf Julia} module takes the linear power spectrum as an input and calculates the non-linear redshift-space power spectrum as a function of $(k,\mu)$ for given bias parameters $b_{{\cal O}}$ as well as the linear growth rate parameter $f={\rm d}\ln D/{\rm d}\ln a$. One can of course calculate the multipole power spectrum as a function of wavenumber $k$ as well.

In this section, we shall compare the outcome of the implementation with the previous results in literature \cite{jeong/komatsu:2006,Jeong2009,Schmittfull:2016,Desjacques:2018} to test the numerical stability and accuracy. 

\subsection{\texorpdfstring{$P_{13}^{\delta\delta}(k)$}{P13delta} and \texorpdfstring{$P_{22}^{\delta\delta}(k)$}{P22delta}
}\label{sec:p13check}
\begin{figure}[ht]
    \centering
    \includegraphics[width=0.75\columnwidth]{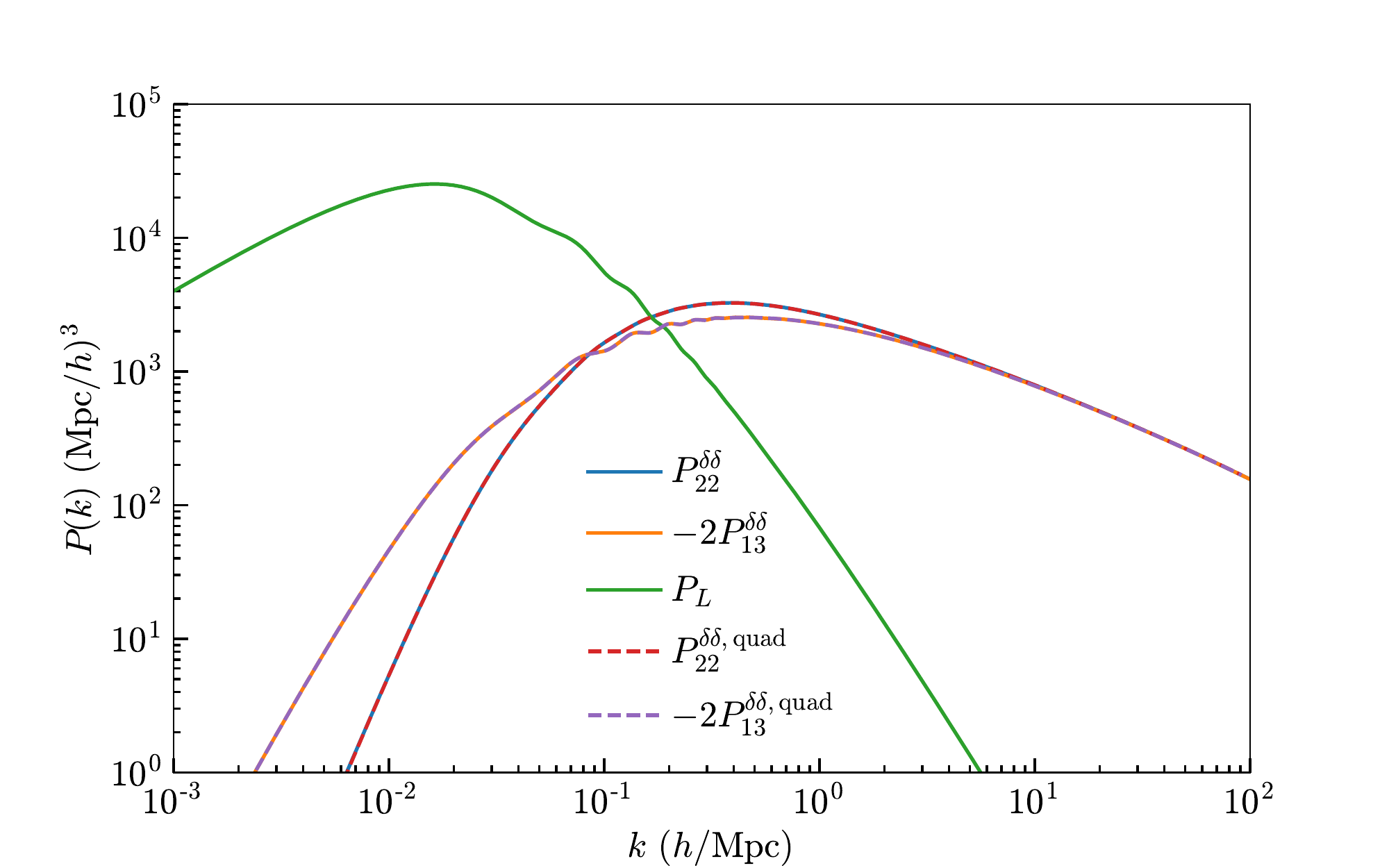}
    \caption{In green is our fiducial power spectrum. In blue is our codes calculation of $P_{22}^{\delta\delta}(k)$, and orange is our calculation of $P_{13}^{\delta\delta}(k)$. Both of these calculations were done using the methods described in \refsec{tests}. The dashed red and purple lines are results of manually integrating equations for $P_{22}^{\delta\delta}(k)$ and $P_{13}^{\delta\delta}(k)$. We see excellent agreement between the two methods, with the FFTLog based method being orders of magnitude faster.}
    \label{fig:svmfig1}
\end{figure}
\begin{figure}
    \centering
    \includegraphics[width=0.49\columnwidth]{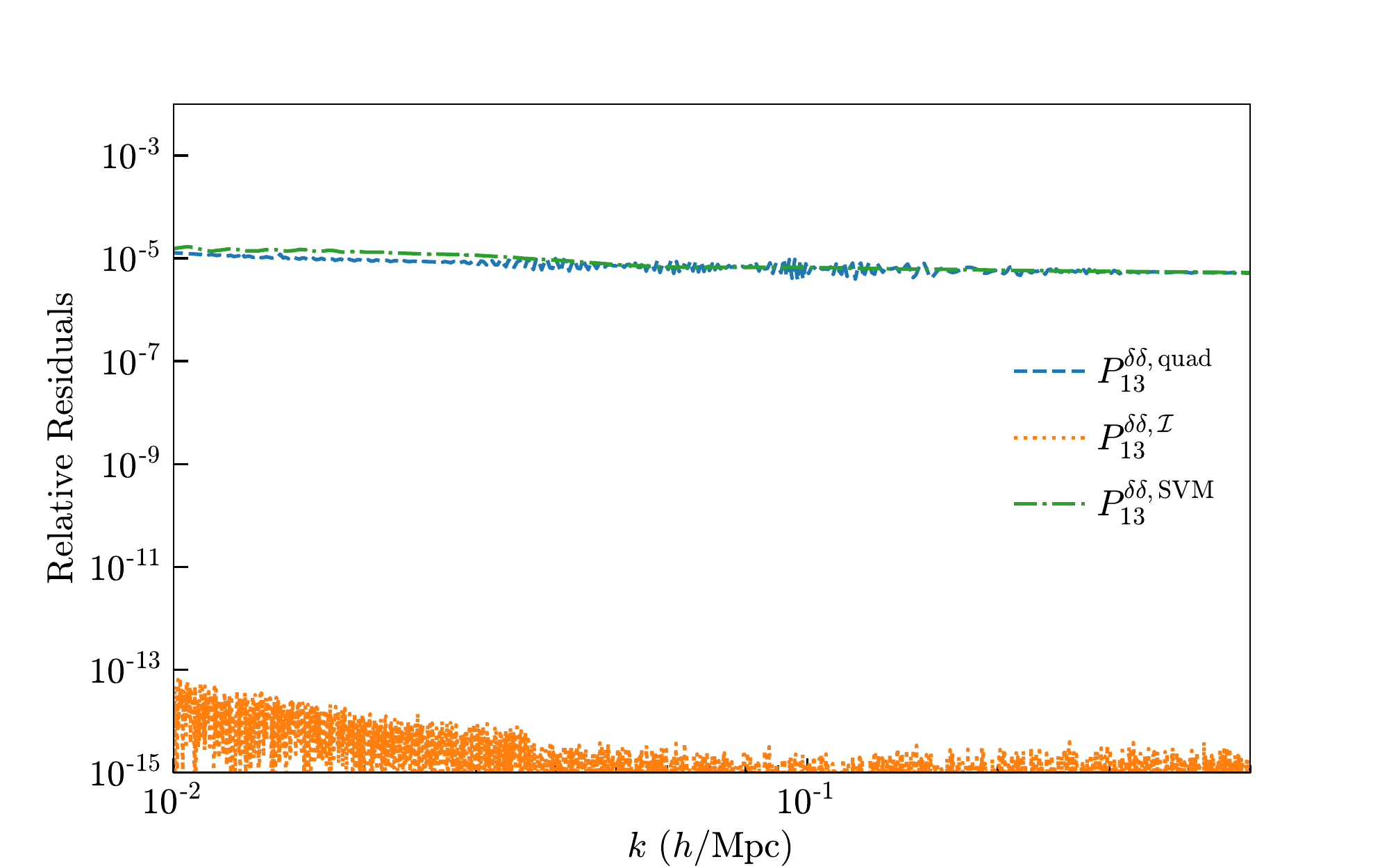}
    \includegraphics[width=0.49\columnwidth]{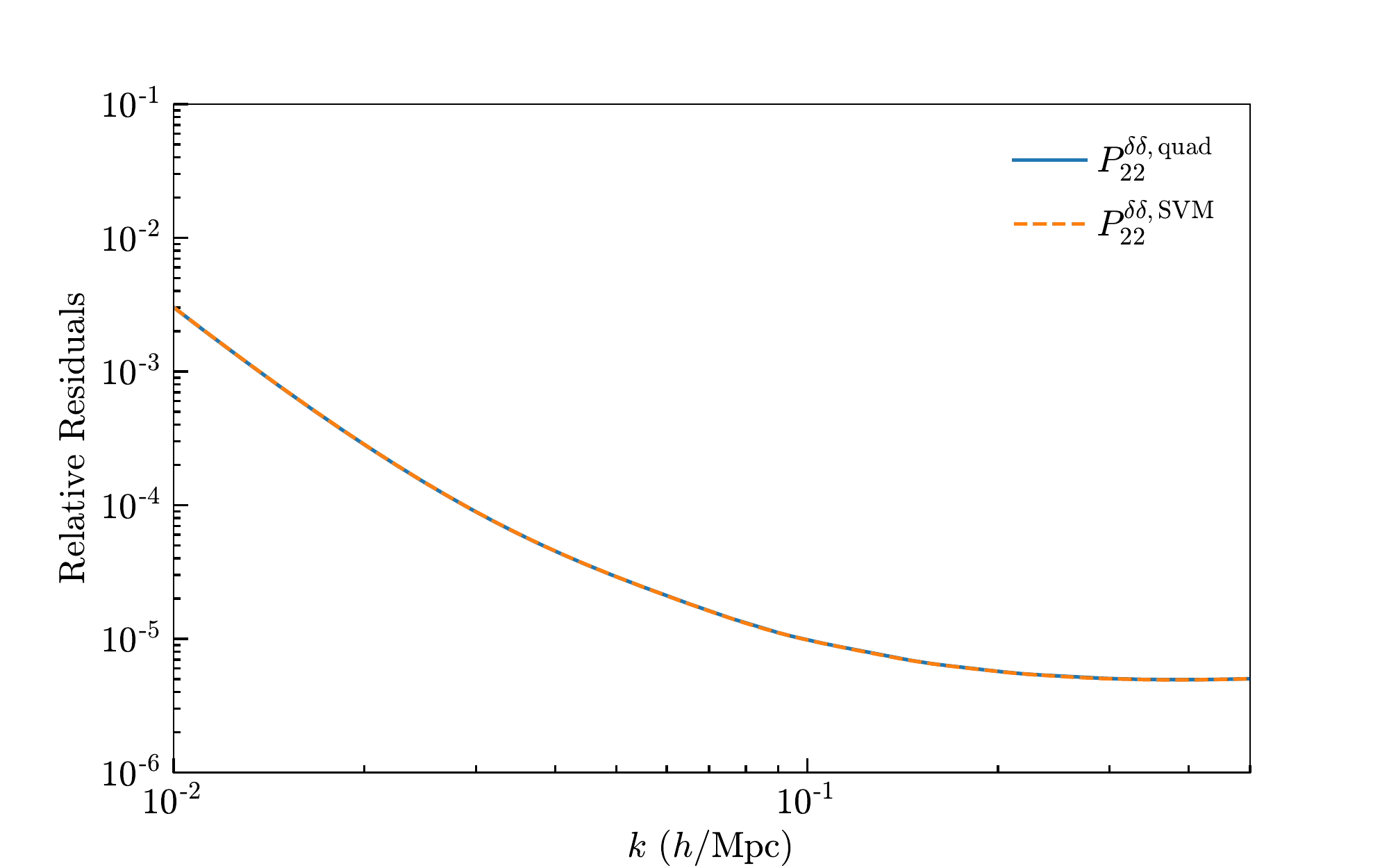}
    \caption{Left: The relative residuals of $P_{13}^{\delta\delta}(k)$ from this work calculated using three different methods. $P_{13}^{\delta\delta}$ is first calculated using our full one-loop code with the only bias parameters being $b_1=1$, then we compare it to the three methods described in \refsec{p13check}. First we compute the integration directly using quadrature, {\sf quadosc}, then as additional tests we use the additional expressions from \refeq{p13I} and \refeq{p13svm}. The errors compared to the quadrature method and \refeq{p13svm} are around $0.001\%$ well within any reasonable bounds, while the error compared to \refeq{p13I} is negligible.
    Right: The relative residuals of $P_{22}^{\delta\delta}(k)$ from this work calculated using two different methods. $P_{22}^{\delta\delta}$ is first calculated using our full one-loop code with the only bias parameter being $b_1=1$, then we compare it to the two methods described in \refsec{p13check}. First we compute the integration directly using quadrature then as an additional test we use \refeq{p22svm}. The error on small scales is consistent with $P_{13}^{\delta\delta}$ at around $0.001\%$ with the large scale errors, where the loop terms are less important, reaching $0.3\%$}
    \label{fig:p1322check}
\end{figure}
First, we perform the comparison with $P_{22}^{\delta\delta}(k)$ and $P_{13}^{\delta\delta}(k)$, both of which can be obtained by setting $b_1=1$ and all other parameters 0, in \reffig{svmfig1}. Note that, in our implementation, $P_{13}^{\delta\delta}$ is given as
\ba \label{eq:p13I}
P_{13}^{\delta\delta, \mathcal{I}} = 3P_L(k)&\(\frac{2}{63}\mathcal{I}_1(k) + \frac{1}{42}\mathcal{I}_2(k)
- \frac{1}{18}\mathcal{I}_3(k) - \frac{1}{18}k^2\sigma_v^2 \),
\ea
with 
\be
\sigma_v^2 = \int_{\bm q} \frac{P_L(q)}{q^2}\,.
\ee
That is, in order to obtain $P_{13}(k)$, in addition to the full 1-3 term with setting $b_O = 0$, $f=0$, $b_1=1$, we need to add $-P_L(k)k^2\sigma_v^2/6$. In \reffig{svmfig1}, we also plot $P_{13}^{\delta\delta,{\rm quad}}(k)$ and $P_{22}^{\delta\delta,{\rm quad}}(k)$ with the same  computation method used in \cite{jeong/komatsu:2006}. As shown there, different calculation methods agree within a sub-percent accuracy for all wavenumbers that we plot here.

We have also calculated $P_{22}^{\delta\delta}(k)$ and $P_{13}^{\delta\delta}(k)$ by using an alternative FFTlog implementation of \cite{Schmittfull:2016}:
\ba \label{eq:p13svm}
P_{13}^{\delta\delta, \rm SVM}(k) = P_L(k) &\( \frac{67}{189}k^2\mathcal{P}_{13}^{0,0}(k) -\frac{1}{3}k^4\mathcal{P}_{13}^{0,-2}(k) + \frac{227}{315}k^3\mathcal{P}_{13}^{1,-1}(k)
- \frac{37}{45}k\mathcal{P}_{13}^{1,1}(k) - \frac{2}{3}k^4\mathcal{P}_{13}^{2,-2}(k) \right.
\vs
&
\left.
- \frac{46}{189}k^2\mathcal{P}_{13}^{2,0}(k) + \frac{76}{105}k^3\mathcal{P}_{13}^{3,-1}(k) + \frac{4}{15}k\mathcal{P}_{13}^{3,1}(k) \) .
\ea
\ba \label{eq:p22svm}
P_{22}^{\delta\delta, \rm SVM}(k) 
=& 4\pi 
\int_0^\infty {\rm d}r\, r^2j_0(kr) 
\left[
\frac{1219}{735}(\xi_0^0(r))^2 + \frac{1}{3}\xi^0_{-2}(r)\xi^0_2(r) - \frac{124}{35}\xi^1_{-1}(r)\xi^1_1(r) + \frac{1342}{1029}(\xi^2_0(r))^2 \right.
\vs
&\qquad\qquad\qquad\qquad\left.+ 
\frac{2}{3}\xi^2_{-2}(r)\xi^2_2(r) -\frac{16}{35}\xi^3_{-1}(r)\xi^3_1(r) + \frac{64}{1715}(\xi^4_0(r))^2\right]\,.
\ea

In \reffig{p1322check}, we plot the residuals between the method developed in this work with the previously discussed methods of calculating $P_{13}^{\delta\delta}$ and $P_{22}^{\delta\delta}$ (quadrature and \refeqs{p13I}{p22svm}). We again find excellent agreement between all methods, with differences consistently below $0.3\%$ validating our numerical implementation. We also get an interesting result from expanding our method analytically for the case of $P_{22}^{\delta\delta}(k)$
\ba
P_{22}^{\delta\delta}(k) = 4\pi \int_0^\infty {\rm d}r\, r^2 j_0(kr) &\( \frac{80}{147} \(\xi_0^0(r)\)^2 - \frac{800}{1029} \(\xi_0^2(r)\)^2 + \frac{80}{343} \(\xi_0^4(r)\)^2 \right.
\vs
&+ k^2 \[\frac{4}{7}\(\xi^3_{-1}(r)\)^2 - \frac{4}{7}\(\xi^1_{-1}(r)\)^2 \] 
\vs
&+ \left. k^4 \[\frac{1}{6}\(\xi^0_{-2}(r)\)^2 + \frac{1}{3}\(\xi^2_{-2}(r)\)^2 \]\)\,,
\ea
which provides a slightly faster way to compute $P_{22}^{\delta\delta}$, going from 12 total transformations to 10.

\subsection{\texorpdfstring{$P_{b2}(k)$}{Pb2} \& \texorpdfstring{$P_{b22}(k)$}{Pb22}}
\begin{figure}
    \centering
    \includegraphics[width=0.49\columnwidth]{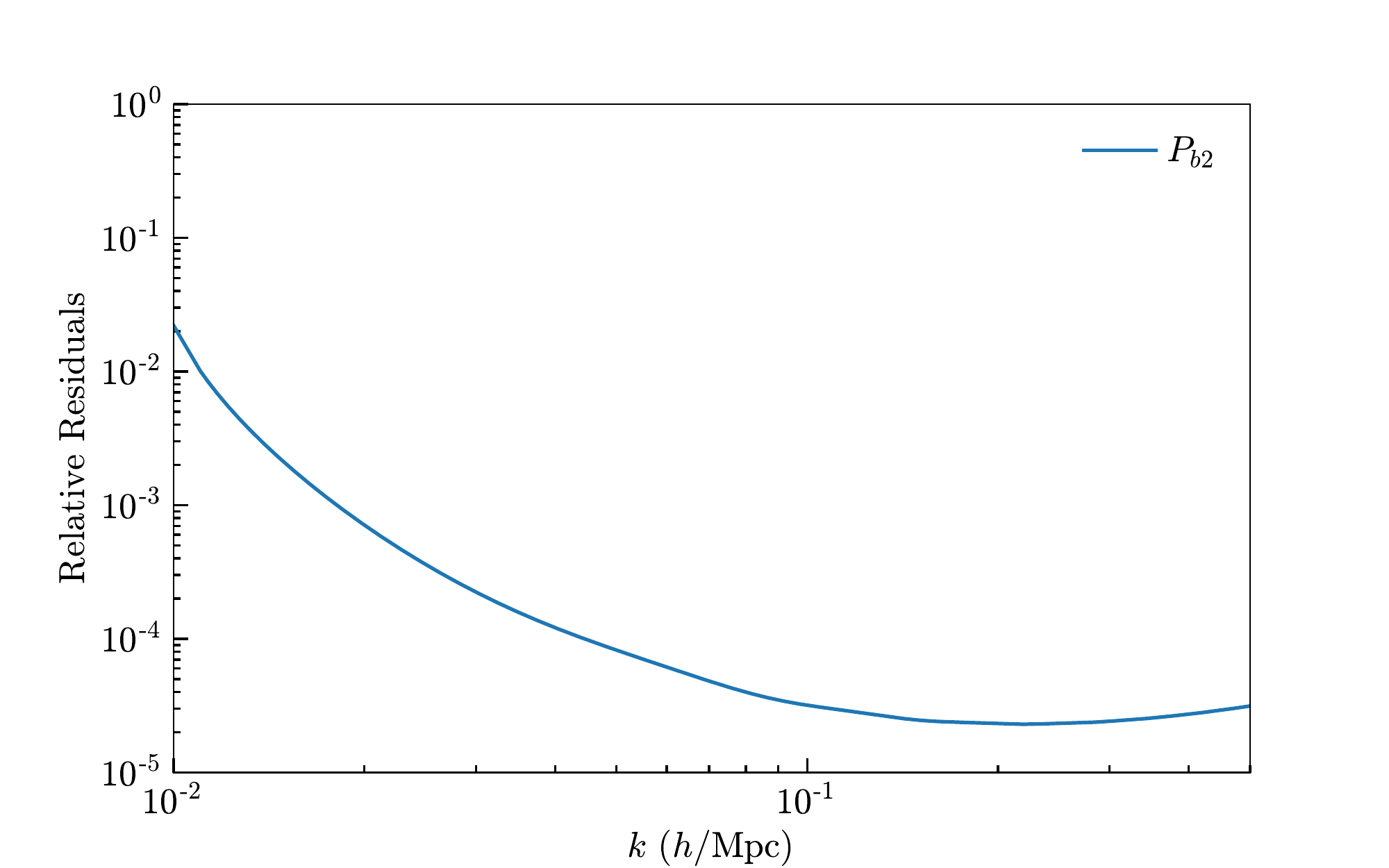}
    \includegraphics[width=0.49\columnwidth]{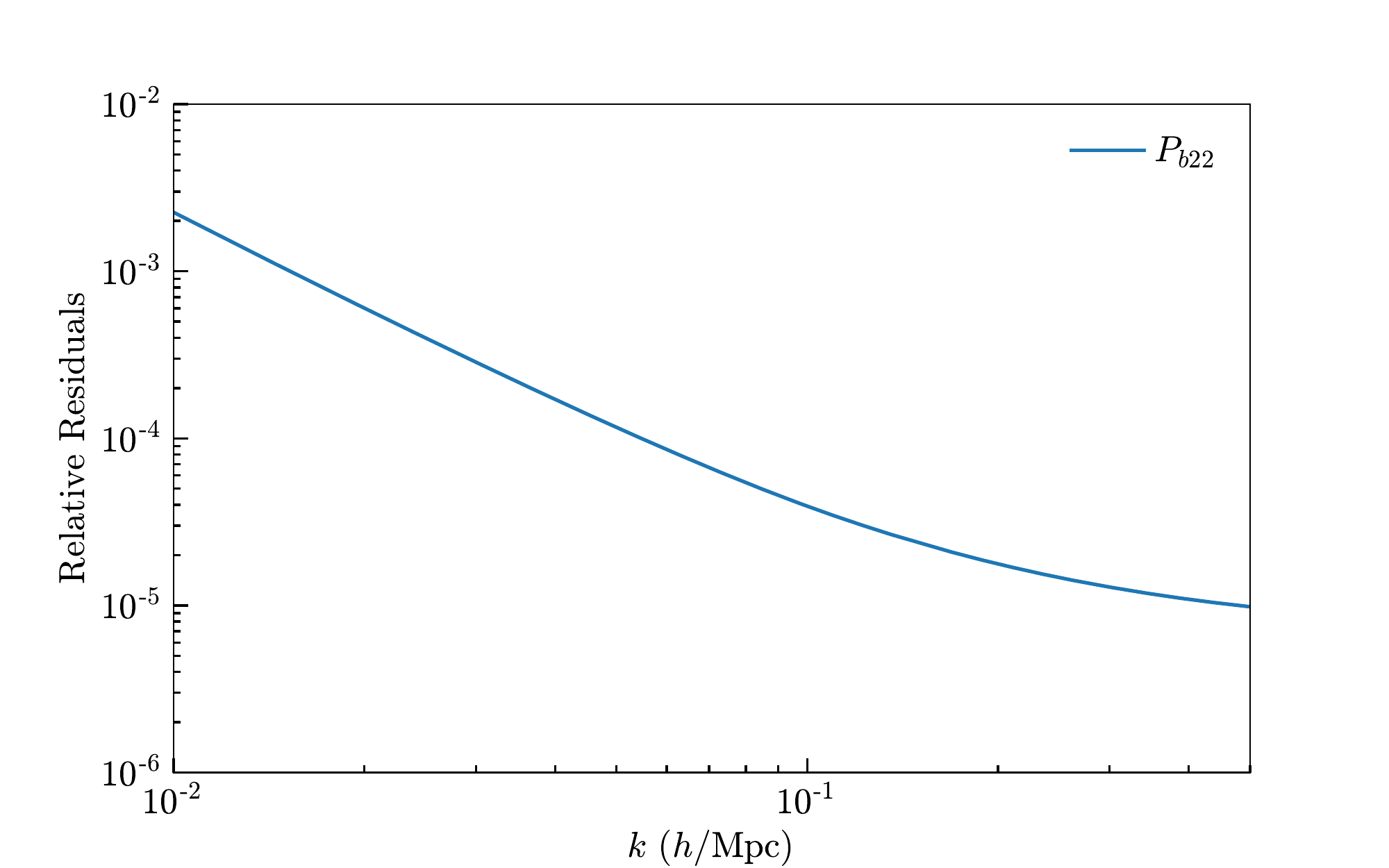}
    \caption{Left: Relative residuals for $P_{b2}$ calculated using the full general bias expansion method described in this work, where we use two different sets of parameters for \refeq{Pb2}, compared to manually integrating using \refeq{pb2dj}. We see a similar trend as in $P_{22}^{\delta\delta}$ with the error maximizing at large scales, where the loop terms are negligible, but remaining under 2\%.
    Right: Relative residuals for $P_{b22}$ calculated using the full general bias expansion method described in this work, where we use two different sets of parameters for \refeq{Pb2}, compared to manually integrating using \refeq{pb22dj}. We see a similar trend to $P_{b2}$ but with generally smaller errors, maxing out around 0.2\%.}
    \label{fig:pb2}
\end{figure}

We next consider two other limiting cases of $P_{22}^{gg,s}(k)$, $P_{b2}(k)$  and $P_{b22}(k)$ which are defined by considering only the local-in-matter-density (LIMD) bias expansion of $P_{22}^{gg,s}(k)$\cite{McDonald2006,Jeong2009}:
\be \label{eq:Pb2}
P_{22}^{\rm LIMD}(k) = b_1^2\[P_{22}^{\delta\delta}(k) + b_2P_{b2}(k) + b_2^2P_{b22}(k) \]\,,
\ee
where the $P_{b2}(k)$ and $P_{b22}(k)$ are defined as
\be\label{eq:pb2dj}
P_{b2}(k) = 2 \int_{\bm q} P_L(q) P_L(|\bm k - \bm q|)F_2^{(s)}(\bm q, \bm k - \bm q)\,,
\ee
and
\be\label{eq:pb22dj}
P_{b22}(k) = \frac{1}{2} \int_{\bm q} P_L(q) \[ P_L(|\bm k - \bm q|) - P_L(q)\]\,.
\ee
On the other hand, we can also extract $P_{b2}(k)$  and $P_{b22}(k)$ using our general bias method by solving the system of linear equation at each wavenumber $k$. In \reffig{pb2}, we show the residuals between the results of the two different implementations. For all wavenubmers for which NLO contributions are relevant, the differences stays within a sub-percent accuracy.

\section{Power Spectrum Response} \label{sec:fisher}
\begin{table}[bt]
\centering
   \begin{tabular}{cccc}
   \hline\hline
\multicolumn{4}{c}{Fiducial parameters for \reffigs{Pkfiducial}{l6fisher}} \\ \hline
\multicolumn{1}{c|}{$b_1$} & \multicolumn{1}{c|}{1.5} & 
\multicolumn{1}{c|}{$b_{\Pi^{[2]}_\parallel}$} &  0\\
\multicolumn{1}{c|}{$b_2$} & \multicolumn{1}{c|}{-0.69} & 
\multicolumn{1}{c|}{$b_{\Pi^{[3]}_\parallel}$} &  0\\
\multicolumn{1}{c|}{$b_\eta$} & \multicolumn{1}{c|}{-1} & 
\multicolumn{1}{c|}{$P^{\{ 0\}}_\epsilon$} &  0\\
\multicolumn{1}{c|}{$b_{K^2}$} & \multicolumn{1}{c|}{-0.14} & 
\multicolumn{1}{c|}{$b_{\nabla^2\delta}$} &  1\\
\multicolumn{1}{c|}{$b_{\delta\eta}$} & \multicolumn{1}{c|}{-1.5} & 
\multicolumn{1}{c|}{$\beta_{\nabla^2 \bm v}$} &  1\\
\multicolumn{1}{c|}{$b_{\eta^2}$} & \multicolumn{1}{c|}{1} &
\multicolumn{1}{c|}{$\beta_{\partial^2_\parallel \bm v}$} & 0 \\
\multicolumn{1}{c|}{$b_{(KK)_\parallel}$} & \multicolumn{1}{c|}{0} & 
\multicolumn{1}{c|}{$P^{\{ 2\}}_\epsilon$} &  0\\
\multicolumn{1}{c|}{$b_{td}$} & \multicolumn{1}{c|}{0.27} & 
\multicolumn{1}{c|}{$P^{\{ 2\}}_{\epsilon\varepsilon_\eta}$} & 0 \\
\multicolumn{1}{c|}{$b_{\delta\Pi^{[2]}_\parallel}$} & \multicolumn{1}{c|}{0} &
\multicolumn{1}{c|}{$b_{(\Pi^{[2]}K)_\parallel}$} &  0\\
\multicolumn{1}{c|}{$b_{\eta\Pi^{[2]}_\parallel}$} & \multicolumn{1}{c|}{0} & 
\multicolumn{1}{c|}{$f$} & 0.53\\
   \hline\hline
\end{tabular}
    \caption{
    The fiducial values of each bias parameters that we take derivatives about for calculating the response function. The fiducial values of $b_1$, $b_2$, and $b_{K^2}$ come from Tab. 6 of \cite{Desjacques:review}, while the fiducial values of $b_\eta$, $b_{\delta\eta}$, $b_{\eta^2}$, $b_{(KK)_\parallel}$, $b_{\Pi^{[2]}_\parallel}$, $b_{\delta\Pi^{[2]}_\parallel}$, $b_{\eta\Pi^{[2]}_\parallel}$, $b_{(\Pi^{[2]}K)_\parallel}$, $b_{\Pi^{[3]}_\parallel}$, $\beta_{\partial^2_\parallel \bm v}$ come from considering no selection effects, see Eq. 2.30 of \cite{Desjacques:2018}. The fiducial higher derivative biases are simply set to 1 arbitrarily. The fiducial value for $b_{td}$ is set by Eq. (2.53) of \cite{Desjacques:review}. The fiducial stochastic parameters are set as 0. Finally the fiducial value of $f$ is set by $f\approx \Omega_m^{0.55}$ \cite{Linder2005} with our fiducial $\Omega_m$ set by \cite{planck2018results}.}
    \label{tab:fiducial}
\end{table}

With selection effects, the expression for the non-linear order one-loop power spectrum contains 22 bias parameters. The consistent cosmological analysis of the galaxy power spectrum in redshift space, therefore, must include these parameters along with the cosmological parameters. Having a plethora of parameters, the natural question is whether any of these parameters are strongly degenerate or not. The answer to this question depends, of course, sensitively on the survey parameters such as survey volume, number density and selection function. We can however glimpse the possible degeneracy between bias parameters by studying the power spectrum response, which is defined as
\be
F_\ell(\theta, k) = \frac{1}{P_L(k)}\left. \frac{{\rm d}P_{{\rm LO + NLO},\ell}^{gg,s}(k) }{{\rm d}\theta} \right|_{\theta=\theta_{f}}\,,
\label{eq:Pkresponse}
\ee
for each parameter $\theta$. Here, $\theta_{f}$ is the fiducial value listed in \reftab{fiducial}. The response appears in the usual statistical analysis based on the Fisher information matrix as following:
\be
F_{ij}
=
\sum_\ell
\sum_k w_\ell(k) 
\left[
\frac{P_L(k)}{P_{{\rm LO + NLO},\ell}^{gg,s}(k)}\right]^2
F_\ell(\theta_i,k) F_\ell(\theta_j,k)\,,
\ee
where 
\be
w_{\ell}(k) \propto \frac{V_{\rm survey} k^2 \delta k}
{\[
1+1/(\bar{n}
P_{{\rm LO + NLO},\ell}^{gg,s}(k)
)\]^2}
\ee 
weights each $k$-mode differently taking into account the cosmic variance (numerator) and the finite galaxy density (denominator) effect. Note that $w_\ell(k)$ is inversely proportional to the variance of the power spectrum multipoles \cite{gebhardt/jeong/etal:2019}. That is, we can think of the $k$-depending response functions as vectors whose inner product is the Fisher information matrix as defined above. The parameter degeneracy happens when the two response functions behave exactly the same way as a function of $k$.

\begin{figure}
    \centering
    \includegraphics[width=\columnwidth]{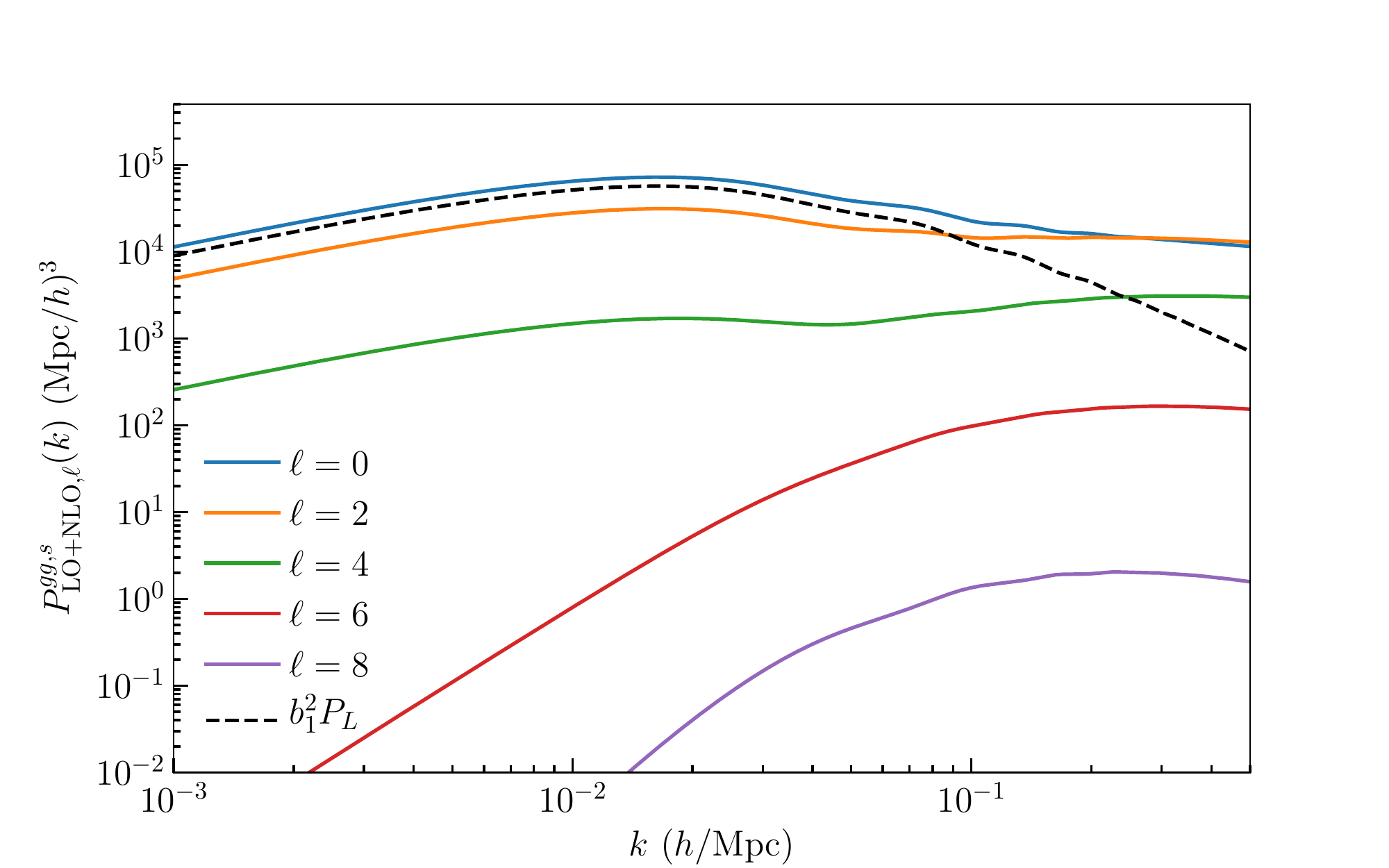}
    \caption{The power spectrum multipoles ($P_{{\rm LO+NLO}\ell}^{gg,s}(k)$) generated using the bias values in \reftab{fiducial}.
    }
    \label{fig:Pkfiducial}
\end{figure}

In this section we use our code for calculating $P_{{\rm LO + NLO},\ell}^{gg,s}$ to examine the power spectrum response function defined in \refeq{Pkresponse} for each bias parameter, and $f$. For reference, we show the power spectrum multipoles with the fiducial parameters shown in \reftab{fiducial} in \reffig{Pkfiducial}. For $\ell<6$, the NLO multipole power spectrum is proportional to $P_L(k)$ on larger scales, which is our motivation of including $P_{L}(k)$ in the definition of the response in \refeq{Pkresponse}.

\begin{figure}
    \centering
    \includegraphics[width=0.49\columnwidth]{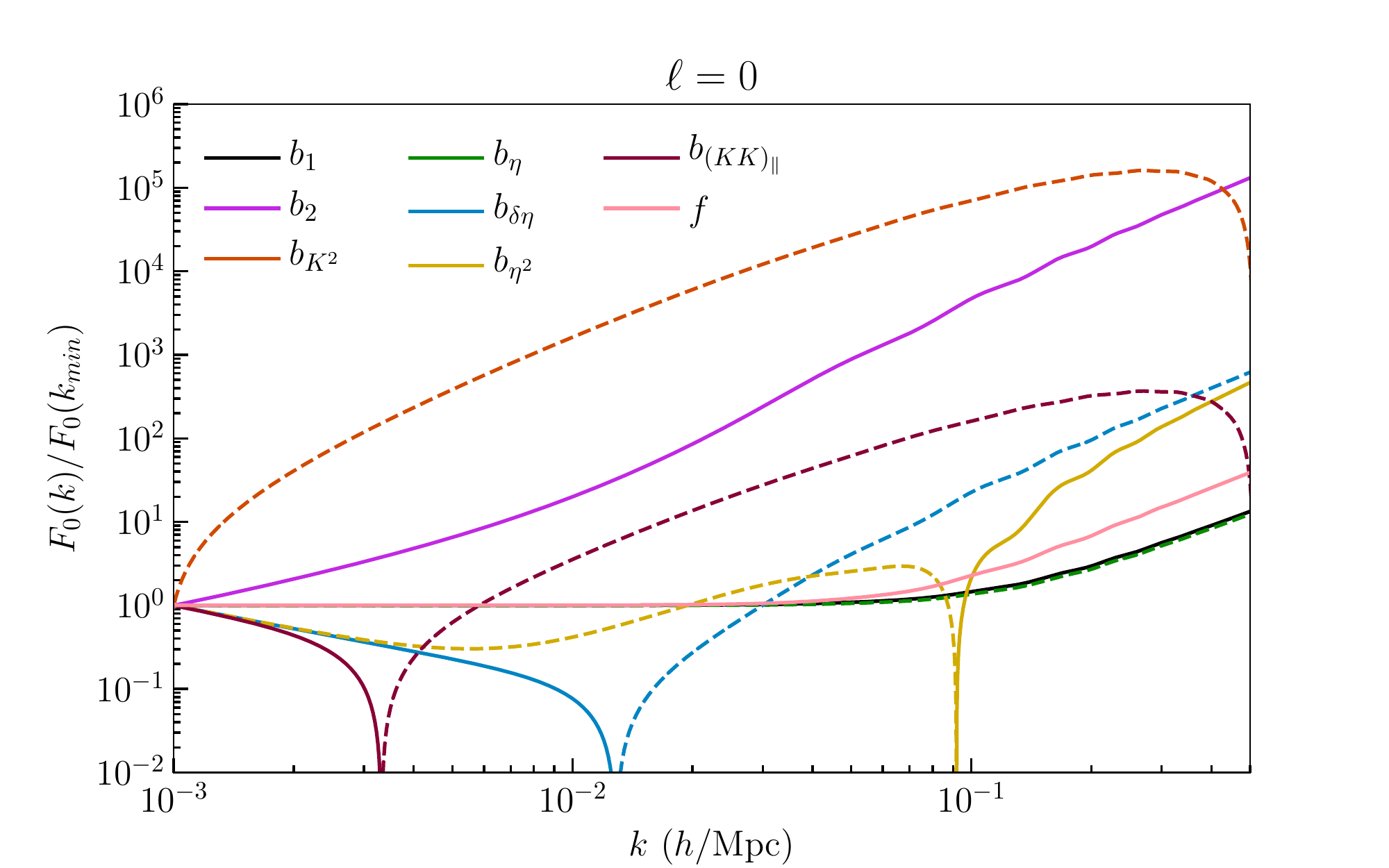}
    \includegraphics[width=0.49\columnwidth]{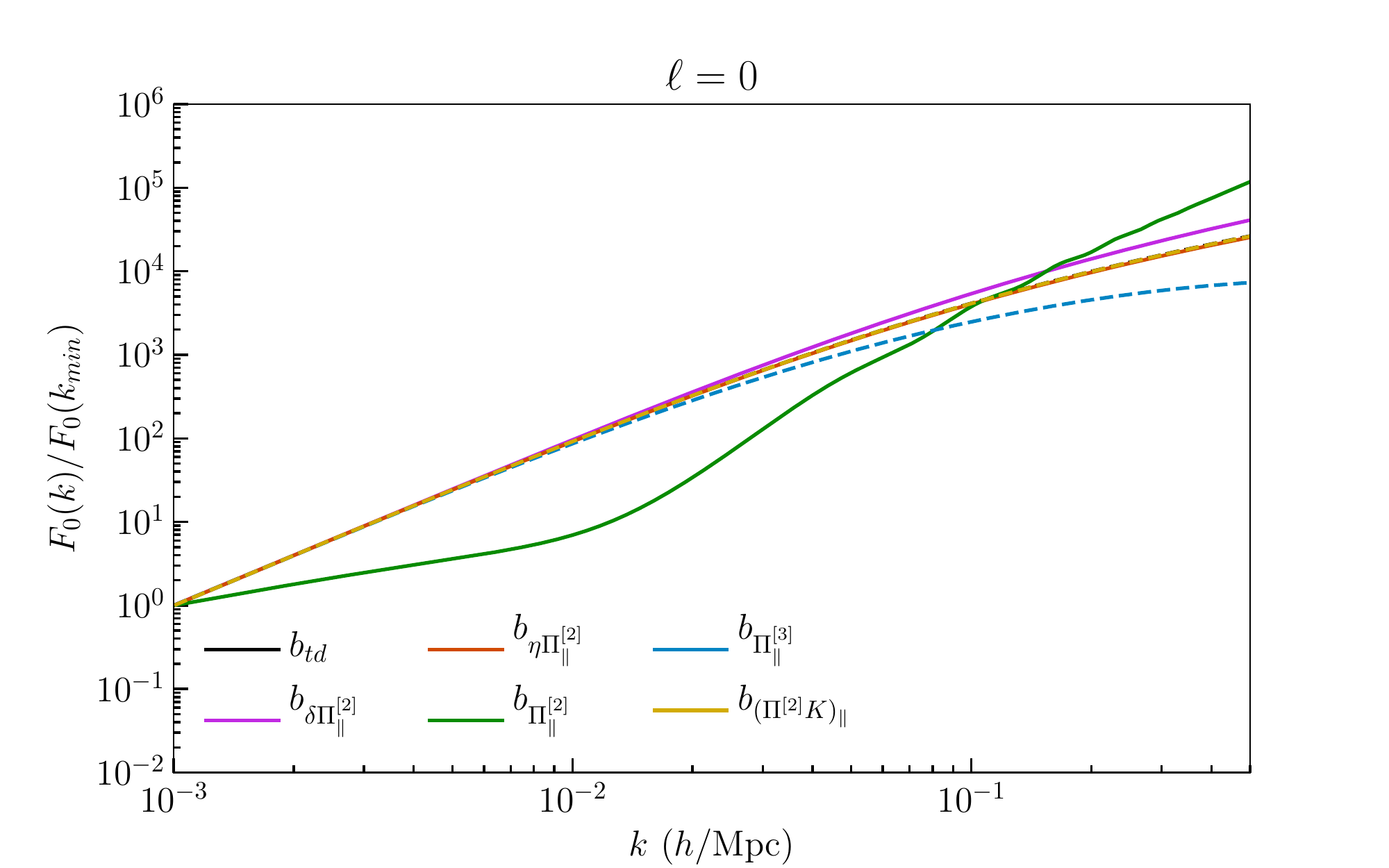}
    \caption{The response function for the monopole LO+NLO power spectrum. We neglect a few bias parameters that are described exactly analytically from \refeq{lhdpole}. For discussion about the degeneracies between parameters see \refsec{fisher}.}
    \label{fig:monofisher}
\end{figure}

In \reffigs{monofisher}{l6fisher}, we show the response for the multipole power spectra $P_\ell(k)$ ($\ell=0$, $2$, $4$, $6$, $8$). The responses for the monopole can be seen in \reffig{monofisher}. While many of the bias parameters are distinct, $b_1$, $b_\eta$, and $f$ are almost perfectly degenerate on large-scales, which is already expected from the linear theory prediction:
$P^{gg,s}_{l+hd,0}(k)\ni (b_1^2 - \frac{2}{3}b_1fb_\eta + \frac{1}{5} f^2b_\eta^2)P_L(k)$.
On small scales, $k\gtrsim0.1\,h/{\rm Mpc}$, however, the NLO contribution potentially distinguishes $f$. We also find that $b_{K^2}$ and $b_{(KK)_\parallel}$ are degenerate on small scales, although they behave differently on large scales. In the right panel, we show that, when just considering the monopole, the parameters $b_{td}$, 
$b_{\delta\Pi^{[2]}_\parallel}$, 
$b_{\eta\Pi^{[2]}}$, and 
$b_{(\Pi^{[2]}K)_\parallel}$ are also nearly perfectly degenerate on large scales.

\begin{figure}
    \centering
    \includegraphics[width=0.49\columnwidth]{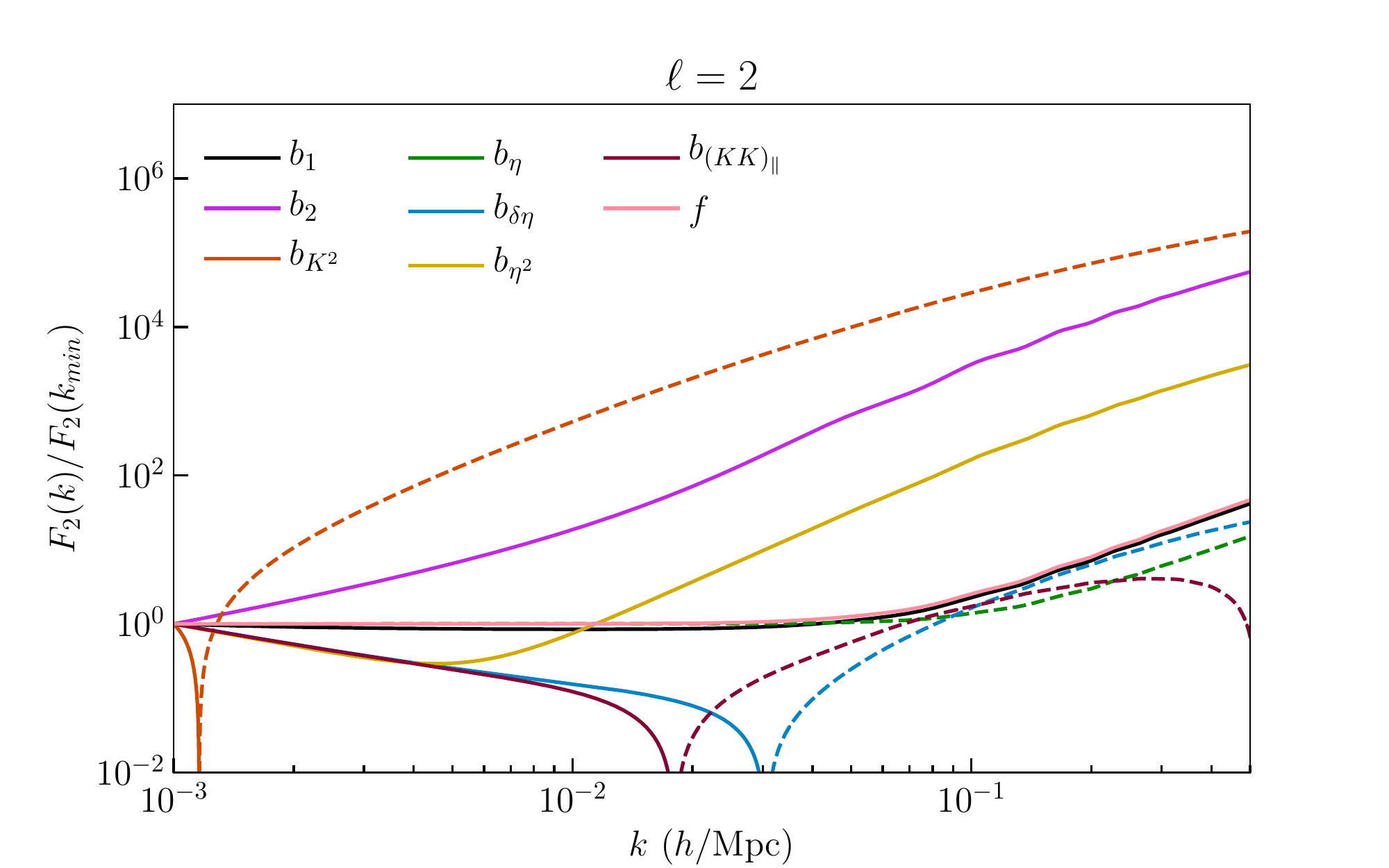}
    \includegraphics[width=0.49\columnwidth]{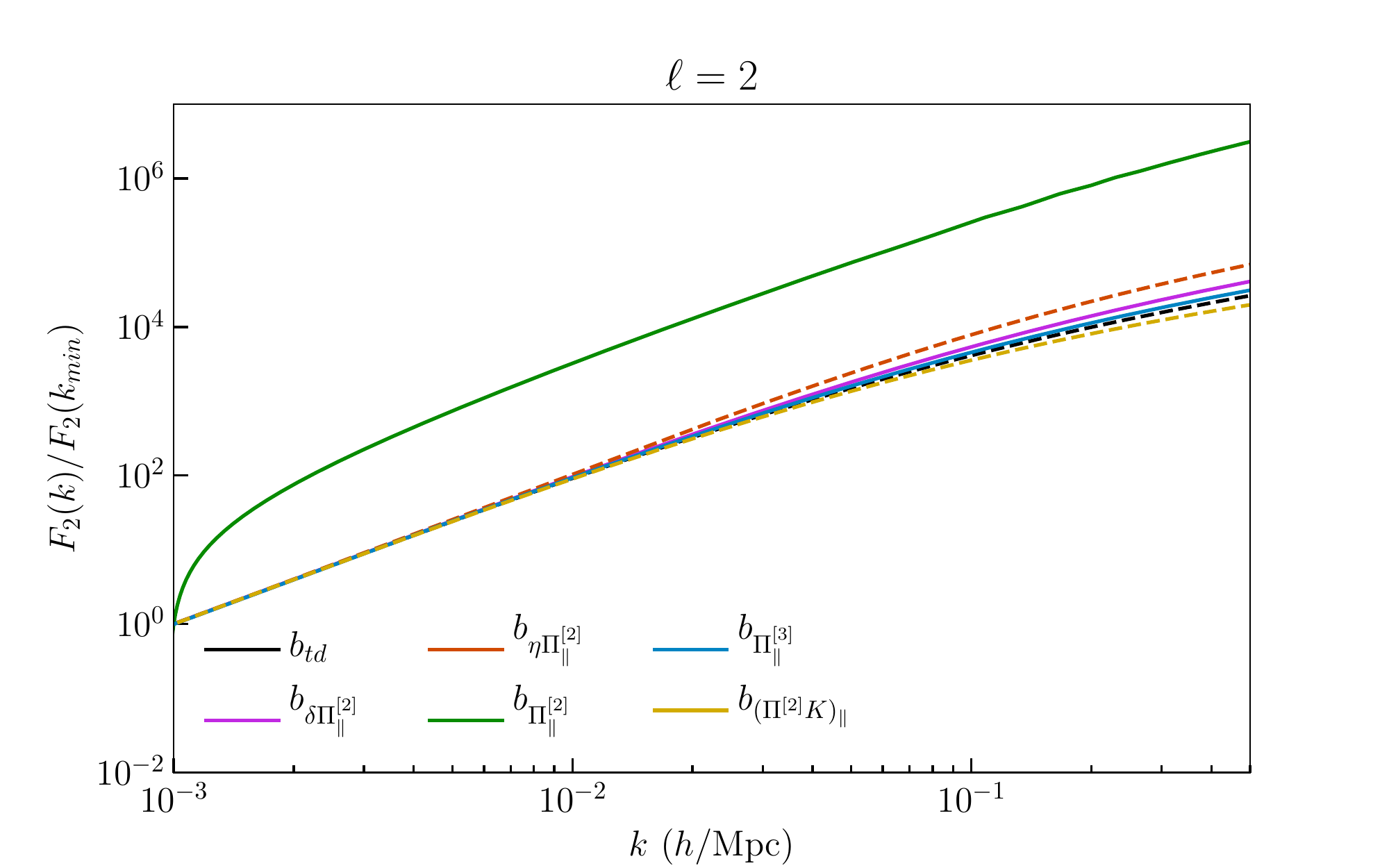}
    \caption{The response function for the quadropole LO+NLO power spectrum. We neglect a few bias parameters that are described exactly analytically from \refeq{lhdpole}. For discussion about the degeneracies between parameters see \refsec{fisher}.}
    \label{fig:l2fisher}
\end{figure}

\begin{figure}
    \centering
    \includegraphics[width=0.49\columnwidth]{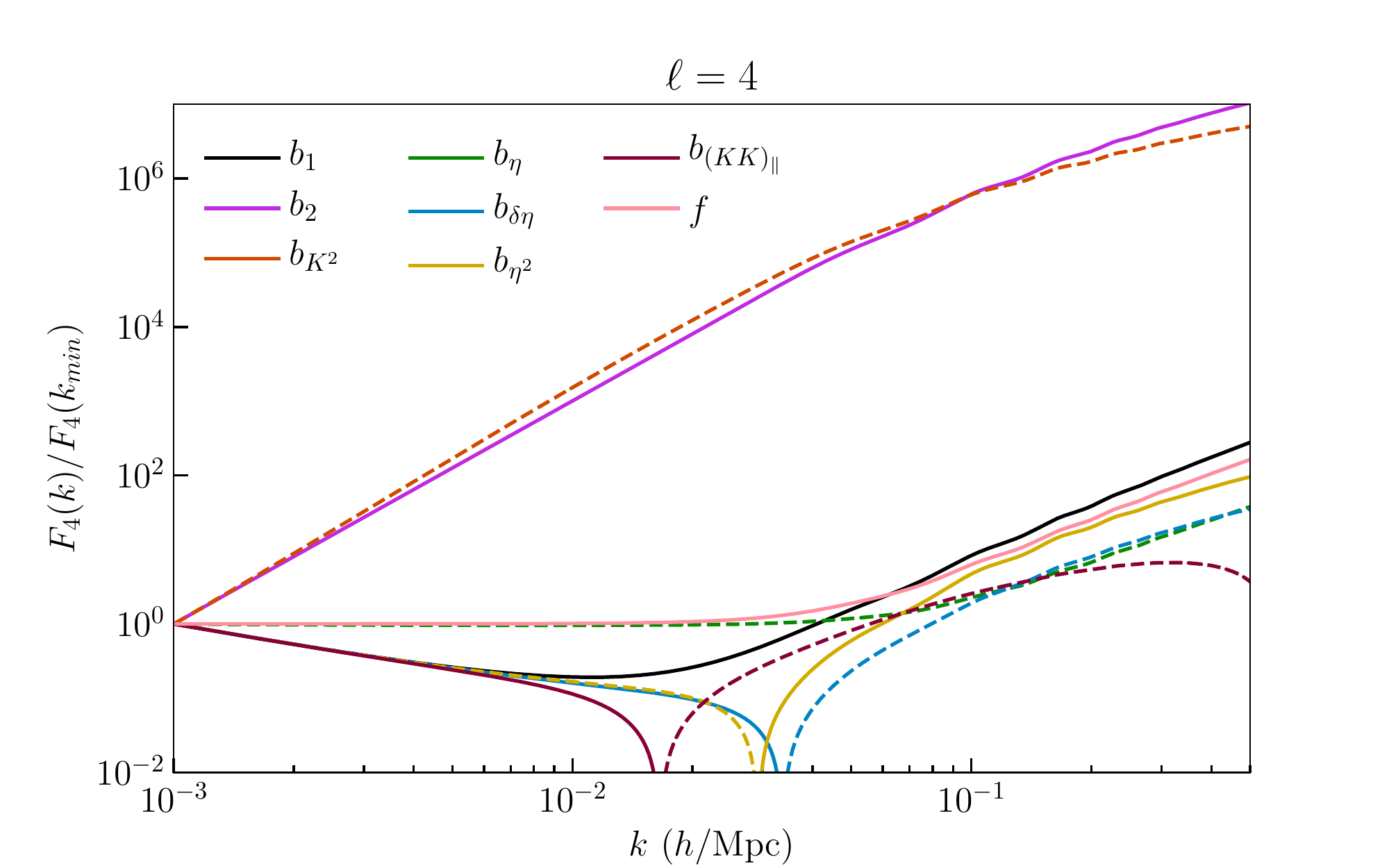}
    \includegraphics[width=0.49\columnwidth]{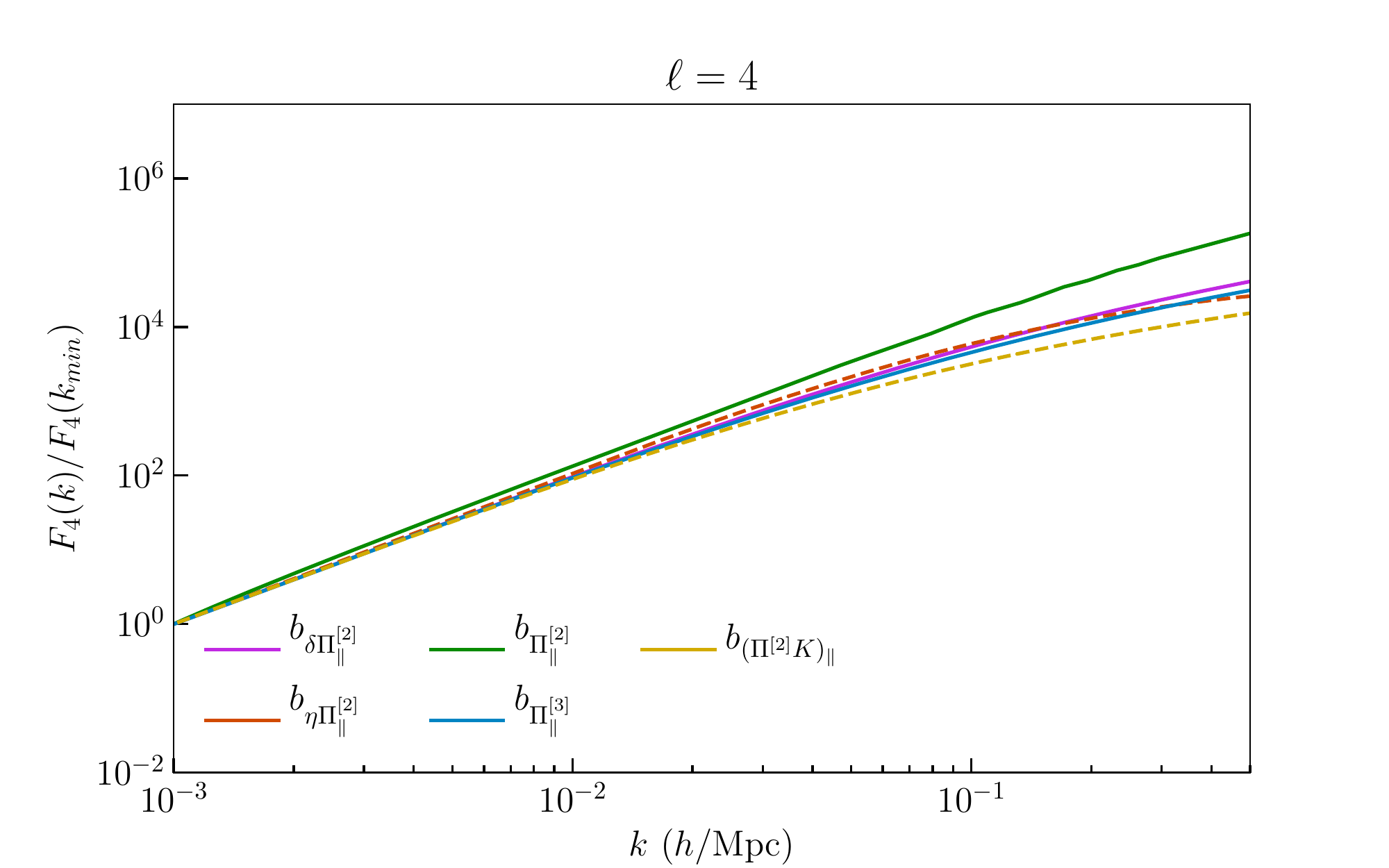}
    \caption{The response function for the octopole LO+NLO power spectrum. We neglect a few bias parameters that are described exactly analytically from \refeq{lhdpole}. For discussion about the degeneracies between parameters see \refsec{fisher}.}
    \label{fig:l4fisher}
\end{figure}

\begin{figure}
    \centering
    \includegraphics[width=0.49\columnwidth]{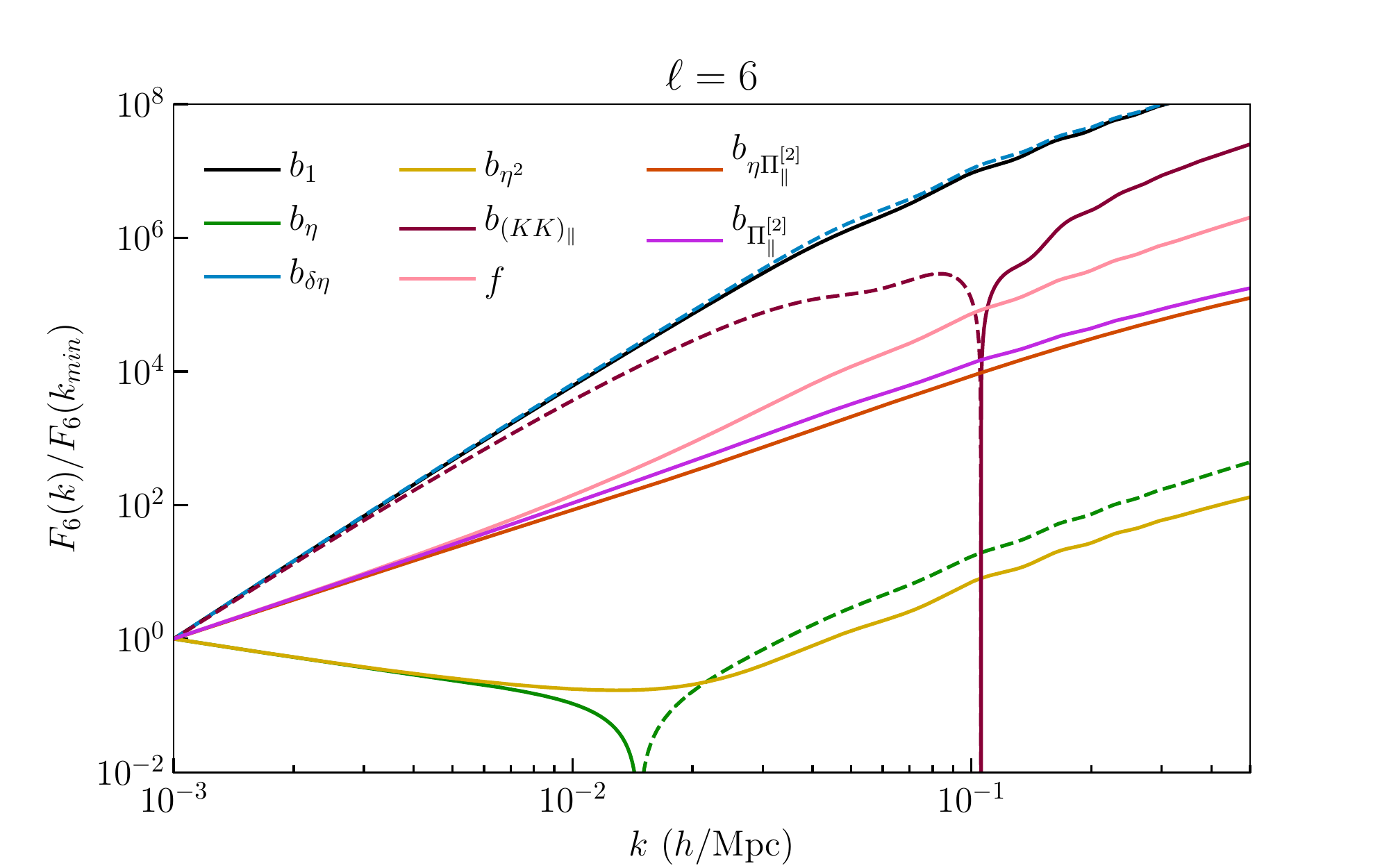}
    \includegraphics[width=0.49\columnwidth]{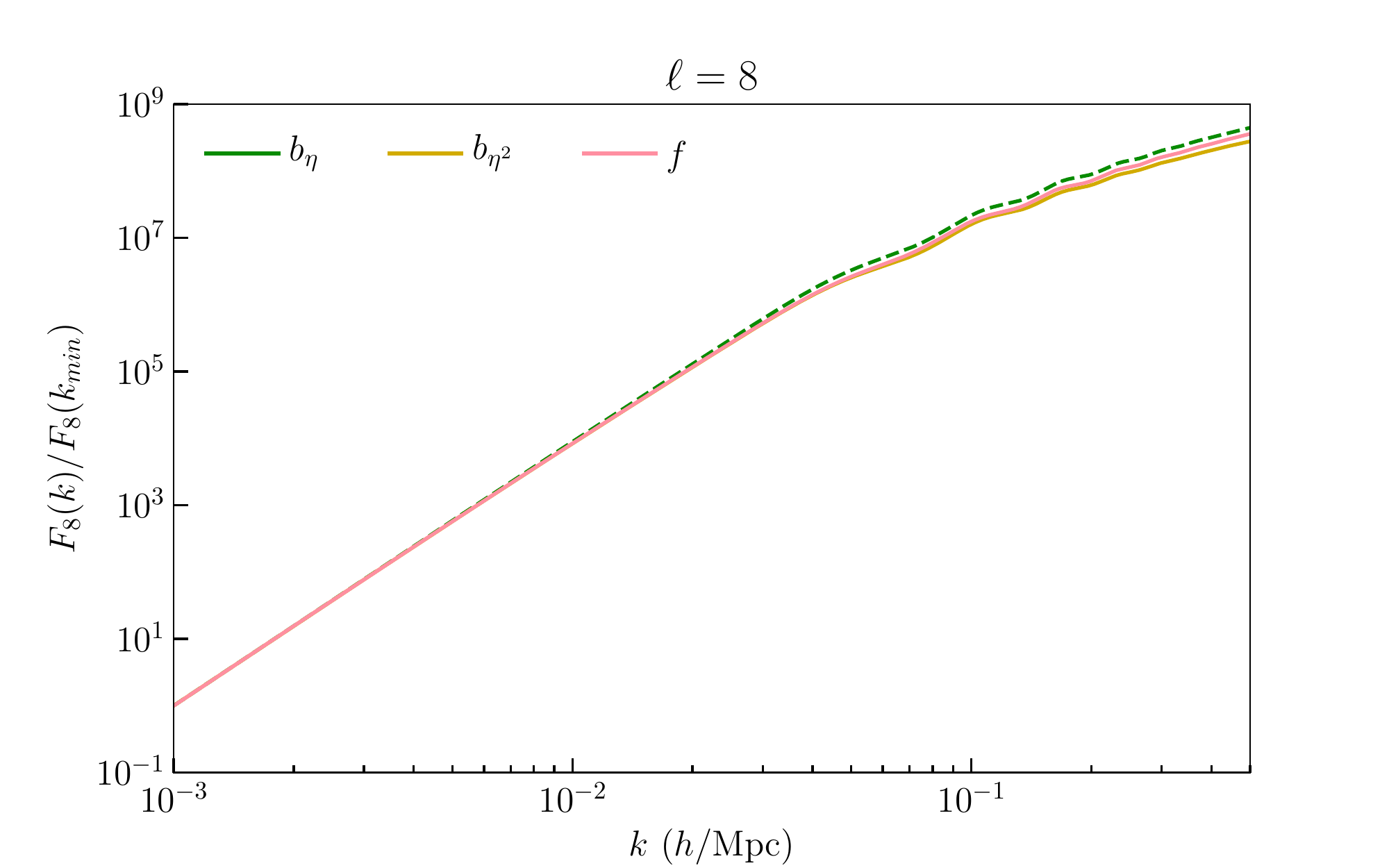}
    \caption{Left: The response function for the hexadecapole LO+NLO power spectrum. We neglect a few bias parameters that are described exactly analytically from \refeq{lhdpole}.
    Right: The response function for the $\ell=8$ LO+NLO power spectrum. For discussion about the degeneracies between parameters see \refsec{fisher}.}
    \label{fig:l6fisher}
\end{figure}

The quadropole ($\ell=2$) responses are presented in \reffig{l2fisher}. Focusing on just the unresolved degeneracies in monopole, we find that, on small scales ($k\gtrsim0.1\,h/{\rm Mpc}$), we gain the ability to distinguish between $b_1$ and $b_\eta$. The quadropole also further breaks the $b_{K^2}$-$b_{(KK)_\parallel}$ degeneracy on small scales. With regards to the four parameters with the worst degeneracy we see some potential for $b_{\eta\Pi^{[2]}_\parallel}$ to be isolated on small scales, leaving only the three parameters $b_{\delta\Pi^{[2]}_\parallel}$, $b_{td}$, and  $b_{(\Pi^{[2]}K)_\parallel}$ which are degenerate with each other. If we further include the octopole ($\ell=4$), \reffig{l4fisher}, then we see that it is independent of $b_{td}$, useful for breaking the primary remaining degeneracy, and that $b_{(\Pi^{[2]}K)_\parallel}$ could potentially be determined based on small scales, leaving no strong degeneracies between the parameters. 
While there is some potential in the $\ell=6$ mode (\reffig{l6fisher}) to clarify some of the parameters, given the small signal-to-noise ratio we anticipate that it does not significantly contribute towards breaking degeneracies, and similarly for the $\ell=8$ mode (\reffig{l6fisher}). 

Of course, the discussion in this section is only based on the shape of the power spectrum response function. We however stress here that the scale- and angular- dependencies of all bias parameters are quite distinctive, so, when applied to the high-$z$ galaxy surveys, the NLO power spectrum has a great potential for exploiting the cosmological information. In particular, the unique scale- and angular-dependence of the linear growth rate parameter $f$ may enable us to measure the parameter as it is, instead of the usual combination of $f\sigma_8$.

\section{Conclusion}\label{sec:conclude}
In this paper we present a fast method of implementing the non-linear galaxy power spectrum in redshift space including the line-of-sight dependent selection bias that arise from, for example, the radiative-transfer effect or tidal alignment effect. 

This work extends the previous fast-integration formalism \cite{Schmittfull:2016,McEwen:2016} using the FFTLog-based method \citep{Siegman:77, TALMAN1978, Hamilton:1999, Gebhardt:2017} leading to multiple orders of magnitude speed up while maintaining accuracy. Including the selection bias parameters, however, we have transformed the loop-integral to ensure convergence.

Our implementation allows the computation of the full NLO power spectrum very quickly, which is essential for the full-shape power spectrum analysis for the future LSS surveys, similar to \cite{Ivanov2019}. This is particularly apt as to unlock the full potential of current and future surveys we need to take into account all possible biases up to 3rd order to reduce modelling uncertainty and extract more information out from the smaller scales.

Although there might be some significant partial degeneracies among the 22 bias parameters that contribute to the NLO galaxy power spectrum, we have demonstrated that the scale- and angular-dependence of the response function for these bias parameters are rather unique. Furthermore, as discussed in \cite{Desjacques:2018}, using the leading order or tree-level bispectrum can help to break many of the degeneracies present due to the direct dependency on the angles between the different $\bm k_i$ and $\nhat$. The methods discussed in this work and that of \cite{Desjacques:review} can also be applied to the NLO bispectrum which we are hopeful will break even more of the degeneracies in these parameters and allow for more precise and unbiased results from future surveys, we leave this for future work.

As discussed in Ref.~\cite{Desjacques:2018}, for many situations some of the selection effects can be argued to be negligible on physical grounds. This of course would greatly improve the cosmological constraints from the analysis. The selection bias may not be negligible for all cases, for example, the radiative transfer effects \cite{Zheng:2011} can be significant for galaxy samples selected based on emission lines like HETDEX \cite{HETDEX:2008}, \textit{WFIRST} \cite{WFIRST:2015}, \textit{Euclid} \cite{Euclid:2011}, and \textit{SPHEREx} \cite{Spherex:2014}. The other primary selection effect is the tidal alignment bias, and although some work has gone into measuring it for early-type galaxies \cite{Martens:2018}, many properties of this effect are still unknown.

One caveat is that the general bias expansion, the higher derivative terms more specifically, introduce an additional length scale that needs to be examined, the non-locality scale of galaxy formation. Any scales smaller than this one cannot be described by a perturbative approach since all higher derivative terms become relevant. For dark matter halos this scale is simply the Lagrangian halo radius, as can be seen in simulations \cite{Lazeyras:2019}, but for galaxies it is unclear what the best answer is. If galaxy formation is entirely controlled by host halos then they have the same non-locality scale, but there are many effects that, if they contribute to galaxy formation, could lead to a larger non-localilty scale. Two significant examples of this are significant radiation field effects, which could have a scale as large as the absorption length for these photons \cite{Arif:1991,Bower:1993,Schmidt:2017}, and cosmic ray heating of the intergalactic medium \cite{Broderick:2012,Lamberts:2015}, which also have large mean free paths.

The {\sf Julia} implementation for the methods described in this paper is available at \url{https://github.com/JosephTomlinson/GeneralBiasPk}.

\acknowledgments

We thank Zvonimir Vlah for providing code samples. This work was supported at Pennsylvania State University by NSF grant (AST-1517363) and NASA ATP program (80NSSC18K1103).

\bibliographystyle{apsrev4-2}
\bibliography{main}

\appendix

\section{\texorpdfstring{$P_{13}^{gg,s}$}{P13ggs} Bias Coefficient Matrices: \texorpdfstring{$\bm{\mathcal{M}}({\cal O})$}{MO}} \label{app:MO}
Here we merely state the matrices. For a derivation see Ref.~\cite{Desjacques:2018} App F.
\ba
\bm{\mathcal{M}}\(O_{td}\) &= \frac{1}{7}
\begin{bmatrix}
4 & -6 & 2 & 0 & 0 \\
0 & 0  & 0 & 0 & 0 \\
0 & 0  & 0 & 0 & 0
\end{bmatrix}
\vs
\bm{\mathcal{M}}\(\delta\Pi^{[2]}_{\parallel}\) &= \frac{1}{7}
\begin{bmatrix}
0 & 0 & 5 & -5 & 0 \\
0 & 0 & -15 & 15 & 0 \\
0 & 0 & 0 & 0 & 0
\end{bmatrix}
\vs
\bm{\mathcal{M}}\(f^{-1}\eta\Pi^{[2]}_{\parallel}\) &= \frac{1}{7}
\begin{bmatrix}
0 & 0 & -\frac{15}{4} & \frac{15}{2} & -\frac{15}{4} \\
-5 & \frac{15}{2} & 20 & -60 & \frac{75}{2} \\
5 & -\frac{15}{2} & -\frac{65}{4} & \frac{125}{2} & - \frac{175}{4}
\end{bmatrix}
\vs
\bm{\mathcal{M}}\((\Pi^{[2]}K)_{\parallel}\) &= \frac{1}{7}
\begin{bmatrix}
\frac{5}{4} & -\frac{15}{8} & \frac{35}{24} & -\frac{5}{6} & 0 \\
\frac{5}{4} & -\frac{15}{8} & -\frac{15}{8} & \frac{5}{2} & 0 \\
0 & 0 & 0 & 0 & 0
\end{bmatrix}
\vs
\bm{\mathcal{M}}\(f^{-1}u^{(2)}_\parallel \partial_\parallel\delta\) &= \frac{1}{7}
\begin{bmatrix}
0 & 0 & 3 & -3 & 0 \\
0 & -3 & -6 & 9 & 0 \\
0 & 0 & 0 & 0 & 0
\end{bmatrix}
\vs
\bm{\mathcal{M}}\(f^{-2}u^{(2)}_\parallel\partial_\parallel\eta\) &= \frac{1}{7}
\begin{bmatrix}
0 & 0 & -\frac{9}{4} & \frac{9}{2} & -\frac{9}{4} \\
-\frac{9}{4} & \frac{27}{8} & \frac{63}{8} & -\frac{63}{2} & \frac{45}{2} \\
\frac{15}{4} & -\frac{21}{8} & -\frac{39}{8} & 30 & - \frac{105}{4}
\end{bmatrix}
\vs
\bm{\mathcal{M}}\(s^k\partial_k\Pi^{[2]}_\parallel\) &= \frac{1}{7}
\begin{bmatrix}
\frac{5}{4} & -\frac{15}{8} & \frac{25}{8} & -\frac{5}{2} & 0 \\
-\frac{15}{4} & \frac{45}{8} & -\frac{75}{8} & \frac{15}{2} & 0 \\
0 & 0 & 0 & 0 & 0 
\end{bmatrix}
\vs
\bm{\mathcal{M}}\(f^{-1}u_\parallel\partial_\parallel\Pi_\parallel^{[2]}\) &= \frac{1}{7}
\begin{bmatrix}
0 & 0 & \frac{15}{4} & - \frac{15}{2} & \frac{15}{4} \\
\frac{15}{4} & - \frac{45}{8} & - \frac{225}{8} & \frac{135}{2} & -\frac{75}{2}\\
-\frac{25}{4} & \frac{75}{8} & \frac{225}{8} & -75 & \frac{175}{4}
\end{bmatrix}
\vs
\bm{\mathcal{M}}\(\Pi^{[3]_\parallel}\) &= \frac{1}{7}
\begin{bmatrix}
\frac{13}{8} & -\frac{39}{16} & \frac{65}{16} & -\frac{13}{4} & 0 \\
-\frac{101}{24} & \frac{101}{16} & -\frac{569}{48} & \frac{39}{4} & 0 \\
0 & 0 & 0 & 0 & 0
\end{bmatrix}
\vs
\bm{\mathcal{M}}\(\delta^{(3)}\) &= \frac{1}{7}
\begin{bmatrix}
\frac{2}{3} & \frac{1}{2} & -\frac{7}{6} & 0 & 0 \\
0 & 0 & 0 & 0 & 0\\
0 & 0 & 0 & 0 & 0
\end{bmatrix}
\vs
\bm{\mathcal{M}}\(f^{-1}\eta^{(3)}\) & = \frac{1}{7}
\begin{bmatrix}
0 & 0 & 0 & 0 & 0 \\
-2 & \frac{3}{2} & \frac{1}{2} & 0 & 0 \\
0 & 0 & 0 & 0 & 0
\end{bmatrix}.
\ea
\\
Note that the above matrix for $f^{-1}\eta^{(3)}$ has had a typo fixed from Ref.~\cite{Desjacques:2018}, replacing $2$ and $-\frac{1}{2}$ with $-2$ and $\frac{1}{2}$ respectively.
We also use
\ba
\bm{\mathcal{M}}\(2KK^{(2)}\) &= \frac{5}{2}\bm{\mathcal{M}}\(O_{td}\) 
\vs
\bm{\mathcal{M}}\(f^{-1}\delta\eta^{(2)}\) &= -\frac{3}{5}\bm{\mathcal{M}}\(\delta\Pi_\parallel^{[2]}\)
\vs
\bm{\mathcal{M}}\(f^{-1}u_\parallel\partial_\parallel\eta^{(2)}\) &= -\frac{3}{5} \bm{\mathcal{M}}\(u_\parallel\partial_\parallel\Pi_\parallel^{[2]}\)
\vs
\bm{\mathcal{M}}\(2f^{-2}\eta\eta^{(2)}\) &= -\frac{6}{5}\bm{\mathcal{M}}\(\eta\Pi_\parallel^{[2]}\)
\vs
\bm{\mathcal{M}}\(2(KK^{(2)})_\parallel\) &= 2\bm{\mathcal{M}}\((\Pi^{[2]}K)_\parallel\).
\ea
\section{New \texorpdfstring{$P_{22}$}{P22} Integrals} \label{app:newint}
Unfortunately the integral forms of $P_{22}^{gg,s}(k,\mu)$ from \cite{Desjacques:2018}, when converted into radial integrals, lead to combinations of divergent integrals that we assume have some canceling divergences but are numerically problematic. To avoid this issue we need to consider a much earlier version of the form of $P_{22}^{gg,s}$ that looks more like
\be
P_{22}^{gg,s}(k,\mu) = 2\int_{\bm q} [Z_2(\bm q, \bm k - \bm q)]^2P_L(q)P_L(|\bm k-\bm q|),
\ee
where $Z_2$ is the bias kernel corresponding to all appropriate parameters, for more discussion of the kernel see \refsec{P22_fftlog}. When considering this form we get integrals like
\begin{widetext}
\be
(2\pi)^3\int_{\bm q} \int_{\bm p} q^{n_1-2}p^{n_2-2}\delta_D(\bm p+\bm q-\bm k)P_L(q)P_L(p)\(\nhat\cdot\qhat \)^a \(\nhat\cdot \phat\)^b \(\phat\cdot\qhat \)^c,
\ee
which we can write as Legendre polynomials like
\be
\mathcal{I}_{a'b'c'}^{n_1n_2} = (2\pi)^3\int_{\bm q}\int_{\bm p} q^{n_1-2}p^{n_2-2}\delta_D(\bm p+\bm q-\bm k)P_L(q)P_L(p)\mathcal{L}_{a'}\(\nhat\cdot\qhat \) \mathcal{L}_{b'}\(\nhat\cdot \phat\) \mathcal{L}_{c'}\(\phat\cdot\qhat \).
\ee
Then expanding the dirac delta into plane waves, \refeq{diracwaves}, and decomposing into angular and radial components
\be
\mathcal{I}_{a'b'c'}^{n_1n_2} = (2\pi)^3\int \frac{dr}{2\pi^2}r^2\int\frac{dq}{2\pi^2}q^2q^{n_1-2}\int \frac{dp}{2\pi^2}p^2p^{n_2-2} P_L(q)P_L(p) I_{a'b'c'},
\ee
where
\be
I_{a'b'c'} = \int \frac{d\Omega_r}{4\pi} \int \frac{d\Omega_q}{4\pi}\int\frac{d\Omega_p}{4\pi} e^{i(\bm p + \bm q - \bm k)\cdot \bm r}\mathcal{L}_{a'}(\nhat\cdot\qhat) \mathcal{L}_{b'}(\nhat\cdot\phat) \mathcal{L}_{c'}(\phat\cdot\qhat),
\ee
is the angular part of the integral. We can write this in spherical harmonic form using \refeq{legtoylm}
\ba
I_{a'b'c'} = \sum\limits_{m_a=-a'}^{a'}
\sum\limits_{m_b}
\sum\limits_{m_c}
&\int \frac{d\Omega_r}{4\pi} 
\int \frac{d\Omega_q}{4\pi}
\int \frac{d\Omega_p}{4\pi} 
e^{-ik\cdot r}  
\(\frac{4\pi}{2a'+1} Y_{a'm_a}(\qhat)Y^*_{a'm_a}(\nhat)e^{iq\cdot r}\) 
\vs
&\times \(\frac{4\pi}{2b'+1}Y_{b'm_b}(\phat)Y^*_{b'm_b}(\nhat)e^{ip\cdot r}\)
\(\frac{4\pi}{2c'+1} Y_{c'm_c}(\phat)Y^*_{c'm_c}(\qhat)\).
\ea
Which we denote
\be
I_{a'b'c'} = \frac{1}{(2a'+1)(2b'+1)(2c'+1)} \mathcal{A}_{a'b'c'},
\ee
to simplify future notation. We then expand some of the exponentials in $\mathcal{A}$ in terms of spherical harmonics using \refeq{exptoylm} and then factor into separate angular parts to get
\ba
\mathcal{A}_{a'b'c'} = (4\pi)^2 &\sum\limits_{m_am_bm_c}\sum\limits_{\ell_a=0}^\infty\sum\limits_{m_{\ell a}=-\ell_a}^{\ell_a}\sum\limits_{\ell_b,m_{\ell b}}
\int d\Omega_r \,
e^{-ik\cdot r} i^{\ell_a+\ell_b}Y^*_{\ell_am_{\ell a}}(\rhat)Y^*_{\ell_bm_{\ell b}}(\rhat)Y^*_{a'm_a}(\nhat)Y^*_{b'm_b}(\nhat) j_{\ell_a}(qr) j_{\ell_b}(pr)
\vs
&\times \int d\Omega_q \,  Y_{a'm_a}(\qhat)Y_{\ell_am_{\ell a}}(\qhat)Y^*_{c'm_c}(\qhat)
\int d\Omega_p \, Y_{b'm_b}(\phat)Y_{\ell_bm_{\ell b}}(\phat)Y_{c'm_c}(\phat).
\ea
Which we can simplify using the Gaunt integral, \refeq{gaunt}, to get
\ba
\mathcal{A}_{a'b'c'} = (4\pi)^2\sum\limits_{m_am_bm_c\ell_a\ell_bm_{\ell a}m_{\ell b}} 
\int d\Omega_r &\, 
e^{-ik\cdot r}(-1)^{m_c}i^{\ell_a+\ell_b}Y^*_{\ell_am_{\ell a}}(\rhat)
Y^*_{\ell_bm_{\ell b}}(\rhat)Y^*_{a'm_a}(\nhat)Y^*_{b'm_b}(\nhat)
\vs
&\times\mathcal{G}^{m_a,m_{\ell a},-m_c}_{a'\ell_ac'}\mathcal{G}^{m_b,m_{\ell b},m_c}_{b'\ell_bc'}j_{\ell_a}(qr) j_{\ell_b}(pr).
\ea
Now we decompose the final exponential to get
\ba
\mathcal{A}_{a'b'c'} &= (4\pi)^3\sum\limits_{m_am_bm_c\ell_a\ell_bm_{\ell a}m_{\ell b}}
\sum\limits_{\ell_r=0}^\infty\sum\limits_{m_r=-\ell_r}^{\ell_r}
(-1)^{m_c}i^{\ell_a+\ell_b}
j_{\ell_a}(qr) j_{\ell_b}(pr)
\mathcal{G}^{m_a,m_{\ell a},-m_c}_{a'\ell_ac'}\mathcal{G}^{m_b,m_{\ell b},m_c}_{b'\ell_bc'} Y^*_{a'm_a}(\nhat)Y^*_{b'm_b}(\nhat)
\vs
&\times i^{-\ell_r}j_{\ell_r}(kr)Y^*_{\ell_rm_r}(\khat) \int d\Omega_r Y_{\ell_rm_r}(\rhat)Y^*_{\ell_am_{\ell a}}(\rhat)Y^*_{\ell_bm_{\ell b}}(\rhat),
\ea
which simplifies to
\ba
\mathcal{A}_{a'b'c'} = (4\pi)^3&\sum\limits_{\ell_a\ell_b\ell_r} 
i^{\ell_a+\ell_b-\ell_r} j_{\ell_a}(qr) j_{\ell_b}(pr)j_{\ell_r}(kr)
\sum\limits_{m_am_bm_r}
Y^*_{a'm_a}(\nhat)Y^*_{b'm_b}(\nhat) Y^*_{\ell_rm_r}(\khat)
\vs
&\times \sum\limits_{m_cm_{\ell a}m_{\ell b}}
(-1)^{m_c+m_{\ell a}+m_{\ell b}}
\mathcal{G}^{m_a,m_{\ell a},-m_c}_{a'\ell_ac'}
\mathcal{G}^{m_b,m_{\ell b},m_c}_{b'\ell_bc'} 
\mathcal{G}^{m_r,-m_{\ell a},-m_{\ell b}}_{\ell_r\ell_a\ell_b}.
\ea
Now we look specifically at the sum over the product of Gaunt integrals. Using \refeq{6j}
\ba
&\sum\limits_{m_cm_{\ell a}m_{\ell b}}
(-1)^{m_c+m_{\ell a}+m_{\ell b}}
\mathcal{G}^{m_a,m_{\ell a},-m_c}_{a'\ell_ac'}
\mathcal{G}^{m_b,m_{\ell b},m_c}_{b'\ell_bc'} 
\mathcal{G}^{m_r,-m_{\ell a},-m_{\ell b}}_{\ell_r\ell_a\ell_b}
\vs
&= \frac{(2\ell_a+1)(2\ell_b+1)(2c'+1)\sqrt{(2\ell_r+1)(2a'+1)(2b'+1)}}{(4\pi)^{3/2}}
\tj{a'}{\ell_a}{c'}{0}{0}{0}
\tj{b'}{\ell_b}{c'}{0}{0}{0}
\tj{\ell_r}{\ell_a}{\ell_b}{0}{0}{0}
\vs
&\times \sum\limits_{m_cm_{\ell a}m_{\ell b}}
(-1)^{m_c+m_{\ell a}+m_{\ell b}}
\tj{a'}{\ell_a}{c'}{m_a}{m_{\ell a}}{-m_c}
\tj{b'}{\ell_b}{c'}{m_b}{m_{\ell b}}{m_c}
\tj{\ell_r}{\ell_a}{\ell_b}{m_r}{-m_{\ell a}}{-m_{\ell b}}
\vs
&= \frac{(2\ell_a+1)(2\ell_b+1)(2c'+1)\sqrt{(2\ell_r+1)(2a'+1)(2b'+1)}}{(4\pi)^{3/2}}
\tj{a'}{\ell_a}{c'}{0}{0}{0}
\tj{b'}{\ell_b}{c'}{0}{0}{0}
\tj{\ell_r}{\ell_a}{\ell_b}{0}{0}{0}
\vs
&\times (-1)^{\ell_a+\ell_b+c'} 
\sj{a'}{b'}{\ell_r}{\ell_b}{\ell_a}{c'}
\tj{a'}{b'}{\ell_r}{m_a}{m_b}{m_r}.
\ea
This gives us
\ba
\mathcal{A}_{a'b'c'} &= (4\pi)^3 \sum\limits_{\ell_a\ell_b\ell_r} 
i^{\ell_a+\ell_b-\ell_r} j_{\ell_a}(qr) j_{\ell_b}(pr)j_{\ell_r}(kr)
\frac{(2\ell_a+1)(2\ell_b+1)(2c'+1)\sqrt{(2\ell_r+1)(2a'+1)(2b'+1)}}{(4\pi)^{3/2}}
\vs
&\times 
\tj{a'}{\ell_a}{c'}{0}{0}{0}
\tj{b'}{\ell_b}{c'}{0}{0}{0}
\tj{\ell_r}{\ell_a}{\ell_b}{0}{0}{0}
(-1)^{\ell_a+\ell_b+c'} 
\sj{a'}{b'}{\ell_r}{\ell_b}{\ell_a}{c'}
\vs
&\times 
\sum\limits_{m_am_bm_r}
Y^*_{a'm_a}(\nhat)Y^*_{b'm_b}(\nhat) Y^*_{\ell_rm_r}(\khat)
\tj{a'}{b'}{\ell_r}{m_a}{m_b}{m_r}.
\ea
Now we look at the product of spherical harmonics in $\mathcal{A}$, which we rewrite with \refeq{ylmylm}, and the new 3j symbol from the Gaunt integral sum
\ba
&\sum\limits_{m_am_bm_r} 
Y^*_{a'm_a}(\nhat)Y^*_{b'm_b}(\nhat)Y^*_{\ell_rm_r}(\khat)\tj{a'}{b'}{\ell_r}{m_a}{m_b}{m_r}
\vs
&= \sum\limits_{m_am_bm_r}
(-1)^{m_a+m_b} Y_{a'-m_a}(\nhat)Y_{b'-m_b}(\nhat)Y^*_{\ell_rm_r}(\khat)\tj{a'}{b'}{\ell_r}{m_a}{m_b}{m_r}
\vs
&= \sum\limits_{m_am_bm_r} \sum\limits_{\ell_km_k}
\mathcal{G}^{-m_a-m_b-m_k}_{a'b'\ell_k}
Y_{\ell_km_k}(\nhat) Y^*_{\ell_rm_r}(\khat)\tj{a'}{b'}{\ell_r}{m_a}{m_b}{m_r}
\vs
&= \sum\limits_{m_r\ell_km_k} \sqrt{\frac{(2a'+1)(2b'+1)(2\ell_k+1)}{4\pi}}
\tj{a'}{b'}{\ell_k}{0}{0}{0}
Y_{\ell_km_k}(\nhat)Y^*_{\ell_rm_r}(\khat) (-1)^{a'+b'+\ell_k} 
\vs
&\times
\sum\limits_{m_am_b}
\tj{a'}{b'}{\ell_k}{m_a}{m_b}{m_k} 
\tj{a'}{b'}{\ell_r}{m_a}{m_b}{m_r}
\vs
&= \sum\limits_{m_r} \sqrt{\frac{(2a'+1)(2b'+1)(2\ell_k+1)}{4\pi}}
\tj{a'}{b'}{\ell_k}{0}{0}{0}
Y_{\ell_km_k}(\nhat)Y^*_{\ell_rm_r}(\khat) (-1)^{a'+b'+\ell_k}
\frac{\delta_{\ell_r\ell_k}\delta_{m_km_r}}{2\ell_r+1}
\vs
&= (-1)^{a'+b'+\ell_r} \frac{\sqrt{(2a'+1)(2b'+1)(2\ell_r+1)}}{(4\pi)^{3/2}}
\tj{a'}{b'}{\ell_r}{0}{0}{0}
\sum\limits_{m_r} \frac{4\pi}{2\ell_r+1}Y_{\ell_rm_r}(\nhat)Y^*_{\ell_rm_r}(\khat)
\vs
&= (-1)^{a'+b'+\ell_r} \frac{\sqrt{(2a'+1)(2b'+1)(2\ell_r+1)}}{(4\pi)^{3/2}}
\tj{a'}{b'}{\ell_r}{0}{0}{0}
\mathcal{L}_{\ell_r}(\mu).
\ea
We now combine everything to get the final form of $\mathcal{A}$
\ba
\mathcal{A}_{a'b'c'} = \sum\limits_{\ell_a\ell_b\ell_r} 
&i^{\ell_a+\ell_b-\ell_r}j_{\ell_a}(qr)
j_{\ell_b}(pr)j_{\ell_r}(kr) 
(-1)^{\ell_a+\ell_b+c'+a'+b'+\ell_r}
\vs
& \times
(2\ell_a+1)(2\ell_b+1)(2c'+1)(2\ell_r+1)(2a'+1)(2b'+1)
\vs
& \times 
\tj{a'}{\ell_a}{c'}{0}{0}{0}
\tj{b'}{\ell_b}{c'}{0}{0}{0}
\tj{\ell_r}{\ell_a}{\ell_b}{0}{0}{0}
\tj{a'}{b'}{\ell_r}{0}{0}{0}
\sj{a'}{b'}{\ell_r}{\ell_b}{\ell_a}{c'}
\mathcal{L}_{\ell_r}(\mu),
\ea
which allows us to write out the final expression for $I$
\ba
I_{a'b'c'} = 
\sum\limits_{\ell_a\ell_b\ell_r} 
&i^{\ell_a+\ell_b-\ell_r}
j_{\ell_a}(qr) j_{\ell_b}(pr) j_{\ell_r}(kr)
(-1)^{\ell_a+\ell_b+c'+a'+b'+\ell_r}
(2\ell_a+1)(2\ell_b+1)(2\ell_r+1)
\vs
& \times
\tj{a'}{\ell_a}{c'}{0}{0}{0}
\tj{b'}{\ell_b}{c'}{0}{0}{0}
\tj{\ell_r}{\ell_a}{\ell_b}{0}{0}{0}
\tj{a'}{b'}{\ell_r}{0}{0}{0}
\sj{a'}{b'}{\ell_r}{\ell_b}{\ell_a}{c'}
\mathcal{L}_{\ell_r}(\mu),
\ea
which we can now plug back in to the original integral to arrive at
\ba \label{eq:slowint}
\mathcal{I}_{a'b'c'}^{n_1n_2} =  
&(2\pi)^3(-1)^{a'+b'+c'}\sum\limits_{\ell_a\ell_b\ell_r}
i^{\ell_a+\ell_b-\ell_r} 
(2\ell_a+1)(2\ell_b+1)
(2\ell_r+1) 
\vs
& \times
\tj{a'}{\ell_a}{c'}{0}{0}{0}
\tj{b'}{\ell_b}{c'}{0}{0}{0}
\tj{\ell_r}{\ell_a}{\ell_b}{0}{0}{0}
\tj{a'}{b'}{\ell_r}{0}{0}{0}
\sj{a'}{b'}{\ell_r}{\ell_b}{\ell_a}{c'}
\mathcal{L}_{\ell_r}(\mu)
\mathcal{R}_{n_1,n_2}^{\ell_a,\ell_b,\ell_r},
\ea
\end{widetext}
where
\be
\mathcal{R}_{n_1,n_2}^{\ell_a,\ell_b,\ell_r} = \int \frac{dr}{2\pi^2} r^2 \xi^{\ell_a}_{n_1-2}\xi^{\ell_b}_{n_2-2}j_{\ell_r}(kr).
\ee
Here we have set $(-1)^{\ell_a+\ell_b+c'+a'+b'+\ell_r} = (-1)^{a'+b'+c'}$ because the conditions on a Wigner-3j symbol with all $m=0$ require that the sum of $l$'s is even. For the same reason the term $i^{\ell_a+\ell_b-\ell_r}$ is always real.

While this formulation is correct, we find that it needs an exceedingly large number of FFTs and is relatively unstable numerically. We instead decide to write the above equation in such a way as to minimize the number of total integrals
\begin{widetext}
\be
\mathcal{I}^{n_1n_2}_{a'b'c} = (2\pi)^3(-1)^{a'+b'+c'}\sum\limits_{\ell_r} \mathcal{L}_{\ell_r}(\mu)(2\ell_r+1) \tj{a'}{b'}{\ell_r}{0}{0}{0} \mathfrak{R}^{n_1n_2}_{a'b'c'\ell_r},
\ee
where
\ba
\mathfrak{R}^{n_1n_2}_{a'b'c'\ell_r} =& \int \frac{dr}{2\pi^2}r^2j_{\ell_r}(kr) \sum\limits_{\ell_a\ell_b} i^{\ell_a+\ell_b-\ell_r}(2\ell_a+1)(2\ell_b+1)
\vs
&\times \tj{a'}{\ell_a}{c'}{0}{0}{0} \tj{b'}{\ell_b}{c'}{0}{0}{0} \tj{\ell_r}{\ell_a}{\ell_b}{0}{0}{0} \sj{a'}{b'}{\ell_r}{\ell_b}{\ell_a}{c'} \xi^{\ell_a}_{n_1-2}\xi^{\ell_b}_{n_2-2} .
\ea
This newer form requires many fewer FFTs, around half as many, so is greatly preferred. This is still not fully optimized however, as the multipole form requires even fewer FFTs, see \refsec{P22_fftlog} for the fully optimized form.
\end{widetext}

\section{Mathematical Identities}\label{app:math}
We make use of the following identities, primarily in \refapp{newint}. Most of these expressions come directly from either \cite{Schmittfull:2016} App. C or \cite{Slepian:2018} App. A.
The dirac delta expands into plane waves as
\be\label{eq:diracwaves}
\delta_D(\bm q) = \int_{\bm r} e^{i\bm q \cdot \bm r}.
\ee
To decompose an exponential into spherical harmonics we use
\be\label{eq:exptoylm}
e^{\pm ik\cdot r} = 4\pi \sum\limits_{\ell=0}^\infty\sum_{m=-\ell}^\ell (\pm i)^\ell j_\ell(kr)Y_{\ell m}(\khat)Y^*_{\ell m}(\rhat). 
\ee
To decompose a Legendre polynomial into spherical harmonics we use
\be\label{eq:legtoylm}
\mathcal{L}_\ell(\qhat\cdot\khat) = \frac{4\pi}{2\ell+1}\sum\limits_{m=-\ell}^\ell Y_{\ell m}(\qhat)Y^*_{\ell m}(\khat). 
\ee
The definition of the complex conjugate of a spherical harmonic is
\be\label{eq:ylm*}
Y^*_{\ell m}(\khat) = (-1)^mY_{\ell-m}(\khat). 
\ee
The Gaunt integral is defined as
\ba\label{eq:gaunt}
\mathcal{G}^{m_1m_2m_3}_{l_1l_2l_3} &= \int d\Omega Y_{l_1m_1}(\khat)Y_{l_2m_2}(\khat)Y_{l_3m_3}(\khat) 
\vs
&= \sqrt{\frac{(2l_1+1)(2l_2+1)(2l_3+1)}{4\pi}}
\tj{l_1}{l_2}{l_3}{0}{0}{0} \tj{l_1}{l_2}{l_3}{m_1}{m_2}{m_3}. 
\ea
To combine the product of two spherical harmonics into a single spherical harmonic we use
\be\label{eq:ylmylm}
Y_{l_1m_1}(\khat)Y_{l_2m_2}(\khat) = \sum\limits_{L=|l_1-l_2|}^{l_1+l_2}(-1)^{m_1+m_2}\mathcal{G}^{m_1,m_2,-m_1-m_2}_{l_1l_2L}Y_{Lm_1+m_2}(\khat).
\ee
We use the definition of the Wigner-6j symbol from \citet[Eq.~35.5.23]{NIST:DLMF}
\ba \label{eq:6j}
&(-1)^{l_1+l_2+l_3} \sj{\ell_1}{\ell_2}{\ell_3}{l_1}{l_2}{l_3} \tj{\ell_1}{\ell_2}{\ell_3}{m_1}{m_2}{m_3}
\vs
&= \sum\limits_{m_1'm_2'm_3'}(-1)^{m_1'+m_2'+m_3'} 
\tj{\ell_1}{l_2}{l_3}{m_1}{m_2'}{-m_3'}
\vs
&\times
\tj{l_1}{\ell_2}{l_3}{-m_1'}{m_2}{m_3'}
\tj{l_1}{l_2}{\ell_3}{m_1'}{-m_2'}{m_3},
\ea
and the orthogonality of the Wigner-3j symbols 
\be \label{eq:3jortho}
\sum\limits_{m_1m_2} (2l+1) \tj{l_1}{l_2}{l}{m_1}{m_2}{m} \tj{l_1}{l_2}{l'}{m_1}{m_2}{m'} = \delta_{ll'}\delta_{mm'}
\ee

\section{FFTLog Biasing Parameter Selection}\label{app:qbest}
In this appendix we discuss our expanded selection function for the optimal biasing parameter for the FFTLog algorithm. The general single Bessel function SBT can be written as
\be
g^{n}_\ell(r) = \int\limits_0^\infty \frac{k^2\, {\rm d}k}{2\pi^2}k^n j_\ell(kr)f(k)\,.
\ee
A crucial part of the FFTLog algorithm is to introduce a power law biasing, $(kr)^q$, with biasing parameter $q$ that reduces the aliasing effect. Unfortunately there is no general criterion for selecting this parameter for any input function. Ref.~\cite{Gebhardt:2017} attempts this problem and finds that for the transformation to be well defined formally:
\be\label{eq:qbounds}
{\rm max}(s_{\rm max}+3+n,\, -\ell +.5) < q < {\rm min}(3+s_{\rm min}+n,\, 2)\,,
\ee
where $s_{\rm max}$ is the slope of the input function $f(k)$ evaluated at the upper range of input parameter $k_{\rm max}$. $s_{\rm min}$ is defined similarly as the slope of $f(k)$ at $k_{\rm min}$. These slopes are defined numerically but for all considered input functions are generally quite stable. Ref.~\cite{Gebhardt:2017} also found the optimal $q$ value to be
\be
q_{\rm best} = n - \frac{s_{\rm min} + s_{\rm max}}{2}\,,
\ee
rounded to be inside the range of formal validity.

Unfortunately this selection criterion is only optimal for the input functions considered in that work, the linear power spectrum. In many other applications, including this work, FFTLog transformations need to be performed on more complicated input functions. For example the transformations involving $P_{13}^{gg,s}(k,\mu)$ have input functions like $\xi^n_\ell(r)/r$ and for $P_{22}^{gg,s}(k,\mu)$ we need to transform linear combinations of products of $\xi$ functions. These input functions are more complicated and the $q_{\rm best}$ found in previous work is not sufficient.

We first make a change by relaxing one of the constraints on being formally convergent, allowing $q$ values greater than 2. This doesn't pose much of a problem since even if the integral is not formally convergent it is still well defined due to analytic continuation. The rest of our corrections use the original formula as a base and then add empirically determined corrections based on general properties of the input function. These empiric corrections are derived by finding where the derivative of the transformation is close to zero with respect to $q$. We note that these corrections occur after $q_{\rm best}$ has been rounded to the bounds in \refeq{qbounds}, neglecting the upper bounds of 2.

The first correction is for an input function that has different signs at each end of the input values, so if ${\rm sgn}(f(k_{\rm min})) \neq {\rm sgn}(f(k_{\rm max}))$ then if $f(k_{\rm min}) < 0$ $q = q_{\rm best} + 0.5$ otherwise $q = q_{\rm best} - 0.5$.

The next correction is for monotonically decreasing input functions, defined as $s_{\rm max} <0$ and $s_{\rm min} < -10^{-4}$. The $s_{\rm min}$ bound is not exactly 0 to handle some cases where the function is too flat to need this correction. The value of this correction depends on what the value of $s_{\rm min}$ is, if $s_{\rm min} > -0.3$ then $q = q_{\rm best} - 1.1$ otherwise $q=q_{\rm best} - 0.4$.

An additional correction that we found is for rapidly decreasing functions, defined as $s_{\rm max} < -6.5$ and this imposes the correction $q = q_{\rm best} - 1$

We found that for input functions with large slopes of different signs, defined as $s_{\rm min} > 3.5$ and $s_{\rm max} < -3.5$, the optimal value is $q = q_{\rm best} + 3$.

The final correction that we use is for functions that are flat for small values of the input parameter but have a large negative slope for large values of the input parameter. This is defined as $0<s_{\rm min}<0.5$ and $s_{\rm max} < -3.5$. For this case we find the optimal choice to be $q = q_{\rm best} - 1.2$.

We note that these conditions are not exclusive and if an input function satisfies multiple conditions then both corrections should apply. For example if a function has both a large $s_{\rm max}$ and a large $s_{\rm min}$, so satisfies both the rapid decrease and both slope large conditions, then optimal value would be $q = q_{\rm best} + 2$.
\end{document}